\documentclass[a4paper,fleqn,usenatbib]{mnras}

\usepackage{newtxtext,newtxmath}

\usepackage[T1]{fontenc}
\usepackage{ae,aecompl}

\usepackage{graphicx}	
\usepackage{amsmath}	
\usepackage{amssymb}	

\usepackage{adjustbox} 
\usepackage{lipsum}

\newcommand{\Me}{M$_\oplus$}
\newcommand{\Mjup}{M$_\mathrm{Jup}$}

\newcommand{\ap}{a_\mathrm{p}}

\newcommand{\Mp}{M_\mathrm{p}}

\newcommand{\Md}{M_\mathrm{d}}
\newcommand{\rmin}{r_{\min}}
\newcommand{\rmax}{r_{\max}}
\newcommand{\amin}{a_{\min}}
\newcommand{\amax}{a_{\max}}

\newcommand{\dgap}{\delta_\mathrm{g}}
\newcommand{\siggap}{\sigma_\mathrm{g}}
\newcommand{\wgap}{w_\mathrm{g}}
\newcommand{\rgap}{r_\mathrm{g}}
\newcommand{\ed}{e_\mathrm{d}}
\newcommand{\alphamm}{\alpha_\mathrm{mm}}

\newcommand{\Mc}{M_\mathrm{c}}
\newcommand{\rc}{r_\mathrm{c}}
\newcommand{\deltarc}{\Delta r_\mathrm{c}}
\newcommand{\sigphi}{\sigma_{\phi}}
\newcommand{\omegac}{\omega_\mathrm{c}}

\title[ALMA observations of HD~107146]{A gap in the planetesimal disc
  around HD~107146 and asymmetric warm dust emission revealed by ALMA}

\author[S. Marino et al.]{ S. Marino,$^{1}$\thanks{E-mail:
    sebastian.marino.estay@gmail.com}, J. Carpenter,$^{2}$
  M. C. Wyatt$^{1}$, M. Booth$^{3}$, S. Casassus$^{4,5}$, V. Faramaz$^{6}$,
  \newauthor{V. Guzman$^{2}$, A. M. Hughes$^{7}$,  A. Isella$^{8}$, G. M. Kennedy$^{9}$,
    L. Matr{\`a}$^{10}$, L. Ricci$^{11}$} \newauthor{ and S. Corder$^{2,12}$.}  \\
  $^{1}$Institute of Astronomy, University of Cambridge, Madingley Road, Cambridge CB3 0HA, UK\\
  $^{2}$Joint ALMA Observatory (JAO), Alonso de Cordova 3107 Vitacura - Santiago de Chile, Chile\\
  $^{3}$Astrophysikalisches Institut und Universit\"atssternwarte, Friedrich-Schiller-Universit\"at Jena, Schillerg\"a\ss{}chen 2-3, 07745 Jena, Germany\\
  $^{4}$Departamento de Astronomía, Universidad de Chile, Casilla 36-D, Santiago, Chile\\
  $^{5}$Millennium Nucleus “Protoplanetary Disks”, Chile\\
  $^{6}$Jet Propulsion Laboratory, California Institute of Technology, 4800 Oak Grove drive, Pasadena CA 91109, USA.\\
  $^{7}$Department of Astronomy, Van Vleck Observatory, Wesleyan University, 96 Foss Hill Drive, Middletown, CT 06459, USA\\
  $^{8}$Department of Physics and Astronomy, Rice University, 6100 Main Street, MS-108, Houston, Texas 77005, USA\\
  $^{9}$Department of Physics, University of Warwick, Gibbet Hill Road, Coventry, CV4 7AL, UK\\
  $^{10}$Harvard-Smithsonian Center for Astrophysics, 60 Garden Street, Cambridge, MA 02138, USA\\
  $^{11}$Department of Physics and Astronomy, California State University Northridge, 18111 Nordhoff Street, Northridge, CA 91130, USA\\
  $^{12}$National Radio Astronomy Observatory, 520 Edgemont Road, Charlottesville, VA 22903-2475, USA\\
}

\date{Accepted XXX. Received YYY; in original form ZZZ}

\pubyear{2015}

\begin{document}
\label{firstpage}
\pagerange{\pageref{firstpage}--\pageref{lastpage}}
\maketitle

\begin{abstract}

While detecting low mass exoplanets at tens of au is beyond current
instrumentation, debris discs provide a unique opportunity to study
the outer regions of planetary systems. Here we report new ALMA
observations of the 80-200~Myr old Solar analogue HD~107146 that
reveal the radial structure of its exo-Kuiper belt at wavelengths of
1.1 and 0.86 mm. We find that the planetesimal disc is broad,
extending from 40 to 140~au, and it is characterised by a circular gap
extending from 60 to 100 au in which the continuum emission drops by
about 50\%. We also report the non-detection of the CO J=3-2 emission
line, confirming that there is not enough gas to affect the dust
distribution. To date, HD~107146 is the only gas-poor system showing
multiple rings in the distribution of millimeter sized
particles. These rings suggest a similar distribution of the
planetesimals producing small dust grains that could be explained
invoking the presence of one or more perturbing planets. Because the
disk appears axisymmetric, such planets should be on circular
orbits. By comparing N-body simulations with the observed visibilities
we find that to explain the radial extent and depth of the gap, it
would require the presence of multiple low mass planets or a single
planet that migrated through the disc. Interior to HD~107146's
exo-Kuiper belt we find extended emission with a peak at $\sim20$~au
and consistent with the inner warm belt that was previously predicted
based on 22$\mu$m excess as in many other systems. This warm belt is
the first to be imaged, although unexpectedly suggesting that it is
asymmetric. This could be due to a large belt eccentricity or due to
clumpy structure produced by resonant trapping with an additional
inner planet.




\end{abstract}




\section{Introduction}
\label{sec:intro}
While exoplanet campaigns have discovered thousands of close in
planets in the last decade, at separations greater than 10~au it has
only been possible to detect a few gas giants, mainly through direct
imaging \footnote{http://exoplanet.eu/} \citep{Marois2008,
  Lagrange2009betapicb, Rameau2013}. Protoplanetary disc observations,
on the other hand, have shown that enough mass in both dust and gas to
form massive planets resides at large stellocentric distances
\citep[see review by][]{Andrews2015}. In addition, the detection of
cold dusty debris discs at tens of au shows that planetesimals can and
do form at tens and hundreds of au in extrasolar systems
\citep[e.g.][]{Su2006, Hillenbrand2008, Wyatt2008, Carpenter2009,
  Eiroa2013, Absil2013, Matthews2014pp6, Thureau2014, Montesinos2016,
  Hughes2018}, although the exact planetesimal belt formation
mechanism is a matter of debate \citep[e.g.][]{Matra2018mmlaw}.

It is natural then to wonder \emph{how far out can planets form?}  In
situ formation of the imaged distant gas giants is challenging as the
growth timescale of their cores can easily take longer than the
protoplanetary gas-rich phase \citep{Pollack1996, Rafikov2004,
  Levison2010}. Gravitational instability was thought to be the only
potential pathway towards in-situ formation at tens of au
\citep{Boss1997, Boley2009}, but the revisited growth timescale of
embryos through pebble accretion could be fast enough to form ice
giants or the core of gas giants during the disc lifetime
\citep{Johansen2010, Ormel2010, Lambrechts2012, Morbidelli2012pa,
  Bitsch2015, Johansen2017}. Alternatively, the observed giant planets
at tens of au may have formed closer in and evolved to their current
orbits by migrating outward \citep{Crida2009}, as could be the case
for HR~8799 with four gas giants in mean motion resonances
\citep{Marois2008, Marois2010} surrounded by an outer debris disc
\citep{Su2009, Matthews2014hr8799, Booth2016}, or may have been
scattered from closer in onto a highly eccentric orbit
\citep{Ford2008, Chatterjee2008, Juric2008}, as has been suggested for
Fomalhaut~b \citep{Kalas2008, Kalas2013, Faramaz2015}. On the other
hand, after the dispersal of gas and dust, planetesimals could
continue growing to form icy planets at tens of au over 100~Myr
timescales; however, numerical studies show that once a Pluto size
object is formed at 30-150~au within a disc of planetesimals, these
are inevitably stirred, stopping growth and the formation of higher
mass planets through oligarchic growth \citep{Kenyon2002, Kenyon2008,
  Kenyon2010}. Thus, it is not yet clear how far from their stars
planets can form. Moreover, the discovery of vast amounts of gas
(possibly primordial) in systems with low, debris-like levels of dust
\citep[so-called ``hybrid discs'', e.g.][]{Moor2017} has opened the
possibility for long lived gaseous discs that could facilitate the
formation of both ice and gas giant planets at tens of au.

 


Broad debris discs provide a unique tool to investigate planet
formation at tens of au. Planets formed at large radii or evolved onto
a wide orbit should leave an imprint in the parent planetesimal belt,
and thus in the dust distribution around the system. Gaps have been
tentatively identified in a few young debris discs using scattered
light observations suggesting the presence of planets at large orbital
radii clearing their orbits from debris, e.g.  HD~92945
\citep{Golimowski2011} and HD~131835 \citep{Feldt2017}. However,
alternative scenarios without planets that could also reproduce the
observed structure have not been ruled out yet in these systems. For
example, multiple ring structures can arise from gas-dust interactions
if gas and dust densities are similar \citep{Lyra2013Natur,
  Richert2017}, which might explain HD~131835's rings since large
amounts of CO gas (likely primordial origin) have been found in this
system \citep{Moor2017}. Moreover, the double ring structure around
HD~92945 and HD~131835 has only been identified in scattered light
images, tracing small dust grains whose distribution can be highly
affected by radiation forces \citep{Burns1979}, therefore not
necessarily tracing the distribution of planetesimals
\citep[e.g.][]{Wyatt2006}.

Only HD~107146, an $\sim80-200$~Myr old G2V star \citep[][and
  references therein]{Williams2004} at a distance of $27.5\pm0.3$~pc
\citep{Gaiamission, Gaiadr1}, has a double debris ring structure
tentatively identified at longer wavelengths thanks to the Atacama
Large Millimeter/submillimeter Array \citep[ALMA,][]{Ricci2015}. At
these wavelengths observations trace mm-sized dust for which radiation
forces are negligible, therefore indicating that the double ring
structure is imprinted in the planetesimal distribution as
well. Moreover, these observations ruled out the presence of gas at
densities high enough to be responsible for such structure. The debris
disc surrounding HD~107146 was first discovered by its infrared (IR)
excess using IRAS data \citep{Silverstone2000}, but it was not until
recently that the disc was resolved by the \emph{Hubble Space
  Telescope} (HST) in scattered light, revealing a nearly face on disc
with a surface brightness peak at 120~au and extending out to
$\sim160$~au \citep{Ardila2004, Ertel2011, Schneider2014}. Despite
HST's high resolution, limitations in subtracting the stellar emission
or a smooth distribution of small dust likely kept the double ring
structure hidden. Using ALMA's unprecedented sensitivity and
resolution, \cite{Ricci2015} showed that this broad disc extended from
about 30 to 150~au, but that it had a decrease in the dust density at
intermediate radii, which could correspond to a gap produced by a
planet of a few Earth masses clearing its orbit at 80~au through
scattering. Finally, analysis of \emph{Spitzer} spectroscopic and
photometric data revealed the presence of an extra unresolved warm
dust component in the system, at a temperature of $\sim$120~K and thus
inferred to be located between 5-15 au from the star
\citep{Morales2011, Kennedy2014}.

\begin{table*}
  \centering
  \caption{Summary of band 6 and band 7 (12m and ACA)
    observations. The image rms and beam size reported corresponds to
    briggs weighting using robust=0.5.}
  \label{tab:obs}
  \begin{adjustbox}{max width=1.0\textwidth}
    \begin{tabular}{lrccccccc} 
  \hline
  \hline
  Observation & Dates & t$_\mathrm{sci}$  & Image rms  & beam size (PA) & Min and max baselines [m]  &  Flux calibrator & Bandpass calibrator & Phase calibrator \\
                &       &      [min]       & [$\mu$Jy]  &                &  (5th and 95th percentiles)            & & &\\
  \hline
  Band 6 - 12m & 24, 27-30 Apr 2017 & 236.4 & 6.8 & $0\farcs67\times0\farcs66$ ($-2.9\degr$) & 41 and 312 &  J1229+0203 & J1229+0203 & J1215+1654\\
  Band 7 - 12m & 11 Dec 2016        & 48.8  & 30.0 & $0\farcs46\times0\farcs37$ ($21.2\degr$) & 48 and 410 & J1229+0203 & J1229+0203 & J1215+1654 \\
  Band 7 - ACA & 20 Oct and 2 Nov 2016   & 135.1 &  245 &  $4\farcs3\times3\farcs4$ ($-76.4\degr$) & 9 and 44 & Titan      & J1256+0547 & J1224+2122\\
               & 22 Mar and 13 Apr 2017  & &&&&&& \\
  Band 7 - 12m+ACA & - & - & 31.1 & $0\farcs47\times0\farcs38$ ($20.8\degr$) & - & -  & -  & -\\
    \hline
  \end{tabular}
  
  \end{adjustbox}
\end{table*}

Despite the tentative evidence of planets producing these gaps around
their orbits, neither the HD~107146 ALMA observations, nor the
scattered light observations of HD~92945 and HD~131835, ruled out
alternative scenarios in which planets are not orbiting within these
gaps, but similar structure is created in the dust distribution
through different mechanisms. These different scenarios have important
implications for the inferred dynamical history of the system and
planet formation. While a planet formed in situ could explain the data
reasonably well, questions arise regarding how a planet of a few Earth
masses could have formed at such large separations, where coagulation
and planetesimal growth timescales are significantly
longer. Alternative scenarios such as the one suggested by
\cite{Pearce2015doublering} to explain HD~107146's gap, could avoid
these issues. In that scenario a broad gap is produced by secular
interactions between a planetesimal disc and a similar mass planet on
an eccentric orbit, which formed closer in and was scattered out by an
additional massive planet. That scenario also predicts that the
planet's orbit should become nearly circular and the planet would be
located at the inner edge of the disc at the current epoch. The model
also predicts the presence of asymmetries in the disc such as spiral
features that would be detectable in deeper ALMA observations.

A second alternative scenario was proposed by \cite{Golimowski2011} to
explain the double-ring structure around HD~92945 seen in scattered
light. As shown by \cite{Wyatt2003}, planet migration can trap
planetesimals in mean motion resonances; resulting in overdensities
that are stationary in the reference frame co-rotating with the
planet. Small dust released from these trapped planetesimals can exit
the resonances due to radiation pressure, forming a double-ring
structure that could be observable in scattered light
\citep{Wyatt2006}. On the other hand, in this planet migration
scenario the distribution of mm-sized dust should match the
planetesimal distribution, with prominent clumps that could be seen in
millimetre observations.

Finally, secular resonances produced by a single eccentric planet in a
massive gaseous disc \citep{Zheng2017} or by two planets formed
interior to the disc could also explain some of the wide gaps
\citep{Yelvertoninprep}, but possibly also leaving asymmetric
features. Hence debris disc observations at multiple wavelengths can
disentangle these different scenarios, and provide insights into the
dynamical history of the outer regions of planetary systems, testing
the existence and origin of planets at tens of au, which otherwise
would remain invisible. Since radiation forces acting on the smallest
dust grains have a significant effect on their distribution, the
planetary perturbations discussed above can be best studied using ALMA
observations which trace the distribution of large ($\sim$0.1-10mm)
grains for which radiation forces are negligible, and thus follow the
distribution of their parent planetesimals.

In this paper we present new ALMA observations of HD~107146 in both
band 6 and 7 (1.1 and 0.86 mm). These observations resolve the broad
debris disc around this system at higher sensitivity and resolution
than the data presented by \cite{Ricci2015}, which showed tentative
evidence of a gap as commented above. This paper is outlined as
follows. In \S\ref{sec:obs} we present the new ALMA observations of
the dust continuum and line emission of HD~107146. Then in
\S\ref{sec:model} we model the data using both parametric models and
the output of N-body simulations to quantify the disc structure and
assess whether a single planet could explain the observations. In
\S\ref{sec:dis} we discuss our results, the origin of the gap and
implications for planet formation; the total mass of HD~107146's outer
disc; the detection of an inner component that could be warm dust; and
our gas non-detection. Finally the main conclusions of this paper are
summarised in \S\ref{sec:conclusions}.



\section{Observations}
\label{sec:obs}
HD~107146 was observed both in band 7 (0.86mm, project 2016.1.00104.S,
PI: S. Marino) and band 6 (1.1mm, project 2016.1.00195.S, PI:
J. Carpenter). Band 7 observations were carried out between October
and December 2016 (see Table \ref{tab:obs}) both using the 12m array
and the Atacama Compact Array (ACA) to recover small and large scale
structures. The total number of antennas for the 12m array was 42,
with baselines ranging from 48 to 410 m (5th and 95th percentiles),
and between 9 to 11 ACA 7m antennas with baselines ranging from 9 to
44 m. The correlator was set up with two spectral windows centered at
343.13 and 357.04 GHz with 2 GHz bandwidths and 15.625 MHz spectral
resolution, and the other two centered at 345.03 and 355.14 GHz with
1.875 GHz bandwidths and 0.976 MHz spectral resolution. The four
windows are used together to study the dust continuum emission, while
the latter two are also used specifically to search for line emission
from CO and HCN molecules in the disc (see \S\ref{sec:gas}). The
weather varied between the multiple ACA band 7 observations, with
average PWV values of 0.38, 0.72, 0.89 and 1.2 mm. The average PWV
during the single 12m observation was 0.56~mm.

Band 6 observations were carried out in April 2017 (see Table
\ref{tab:obs}) using the 12m array only. We requested observations
using two different 12m antenna configurations to recover well the
large scale structure and, at the same time, achieve high spatial
resolution, but due to time constraints the compact configuration
observations were never carried out. We find however that the
observations with the extended configuration have baselines short
enough to recover the large scale structure (see \S\ref{sec:continuum}
below). The total number of 12m antennas varied between 41 and 42,
with baselines ranging from 41 to 312 m (5th and 95th
percentiles). The correlator was set up with four spectral windows
centered at 253.60, 255.60, 269.61 and 271.61 GHz with 2 GHz
bandwidths and 15.625 MHz spectral resolution. The four are used
together to study the dust continuum emission only. The weather also
varied between the multiple band 6 observations, with PWV values of
0.89, 0.33, 0.30, 0.82 and 1.56~mm. Despite these variations we
decided to use all the data sets to obtain the highest possible
S/N. Calibrations were applied using the pipeline provided by ALMA and
CASA 4.7. The total time on source for band 6 was 236 min, and 184 min
for band 7 (49 and 135 min for the 12m array and ACA,
respectively). Below we present the image analysis of continuum and
line observations.

\subsection{Continuum}
\label{sec:continuum}

 \begin{figure*}
  \centering
 \includegraphics[trim=0.0cm 0.0cm 0.0cm 0.0cm, clip=true, width=0.9\textwidth]{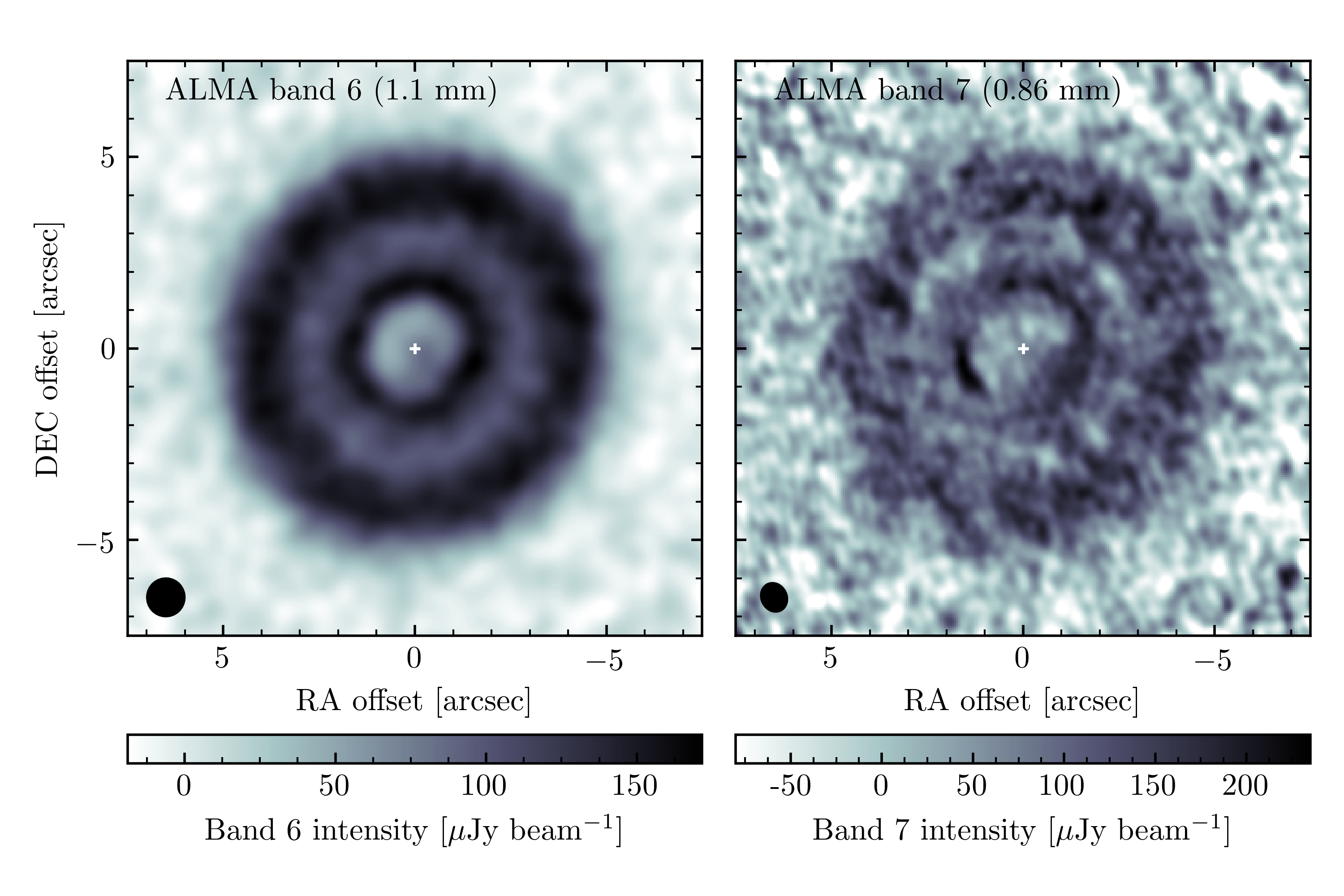}
 \caption{Clean images of HD~107146 at 1.1 (left) and 0.86 mm (right)
   using natural weights and primary beam corrected. The band 7 image
   is obtained after combining the 12m array and ACA data. The stellar
   position is marked with a white cross at the center of the image,
   while the beams of band 6 ($0\farcs80\times0\farcs79$) and band 7
   ($0\farcs58\times0\farcs47$) are represented by black ellipses in
   the bottom left corners. The image rms at the center is 6.3 and 27
   $\mu$Jy~beam$^{-1}$ increasing with distance from the center and
   reaching values of 7.0 and 34 $\mu$Jy~beam$^{-1}$ at $5\arcsec$
   from the center. }
 \label{fig:almab6b7}
\end{figure*}

 \begin{figure}
  \centering
 \includegraphics[trim=0.0cm 0.0cm 0.0cm 0.0cm, clip=true, width=1.0\columnwidth]{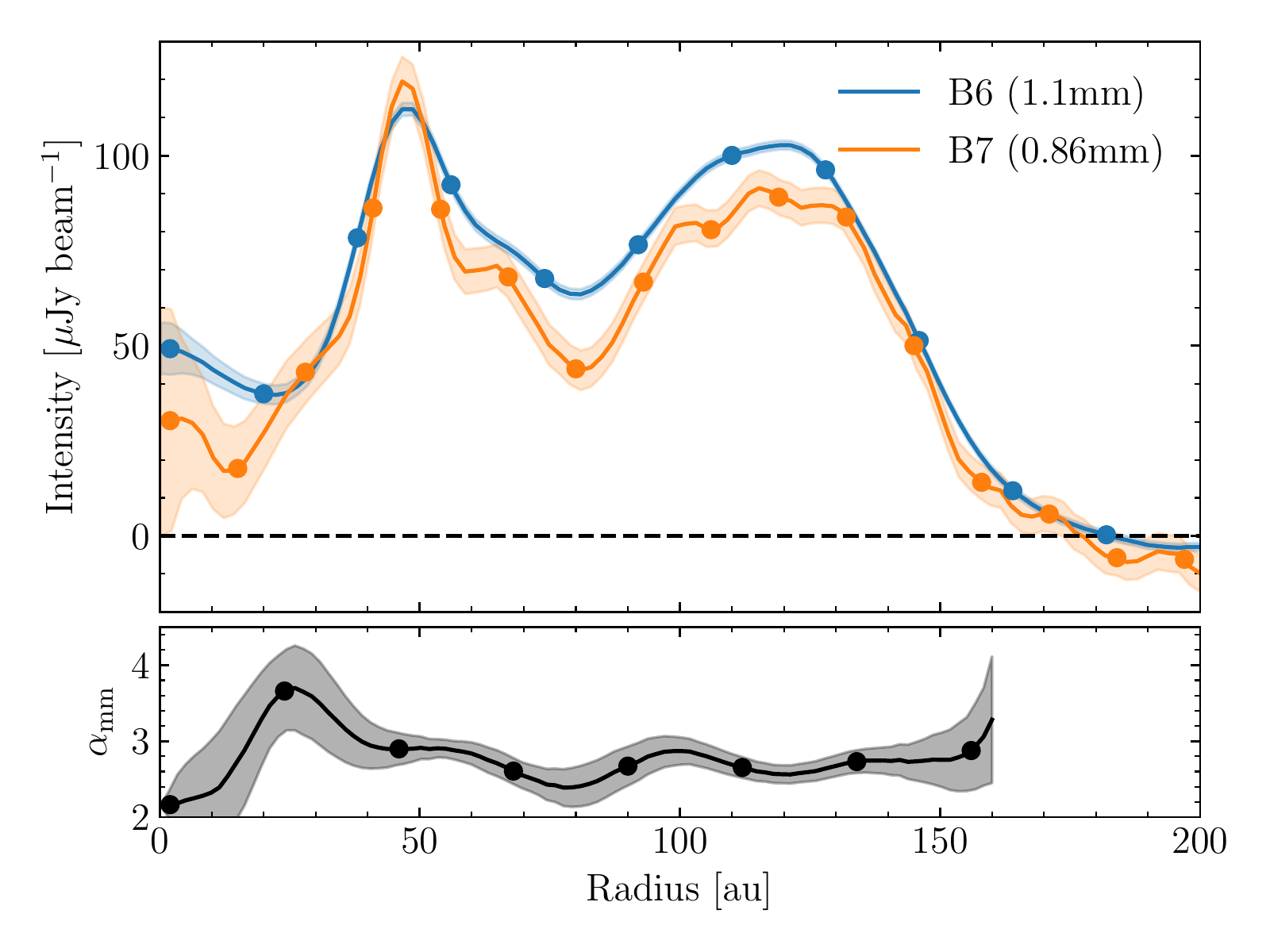}
 \caption{Average intensity (top) and spectral index (bottom) profile
   computed azimuthally averaging the Clean and spectral index maps
   over ellipses oriented as the disc in the sky. The blue and orange
   lines are obtained from the band 6 and band 7 Clean images, using
   Briggs (robust=0.5) weights, while the spectral index is computed
   using natural weights. The shaded areas represent the 68\%
   confidence region over a resolution element (represented by circles
   spaced by 18 au for band 6, 13 au for band 7 and 22 au for the
   spectral index).}
 \label{fig:radial_profile}
\end{figure}

Continuum maps at band 6 and 7 are created using the \textsc{clean}
task in CASA 4.7 \citep{casa} and are presented in Figure
\ref{fig:almab6b7}. We adopt natural weighting for band 6 and 7
(12m+ACA combined) for a higher signal to noise. These maps have a rms
of 6.3 and 27 $\mu$Jy~beam$^{-1}$ at the center, which increases
towards the edges as the maps are corrected by the primary beam,
reaching values of 7.0 and 34 $\mu$Jy~beam$^{-1}$ at $5\arcsec$
(140~au) for band 6 and 7, respectively. The band 6 and band 7
synthesised beams have dimensions of $0\farcs80\times0\farcs79$ and
$0\farcs58\times0\farcs47$\footnote{Note that the rms and beam sizes
  are different to the one reported in Table \ref{tab:obs} as the
  imaging is done with natural weights to increase the S/N},
respectively, reaching an approximate resolution of 22~au for band 6
and 15~au for band 7. The higher resolution and sensitivity of these
new data reveal a nearly axisymmetric broad disc with a large decrease
or gap in the surface brightness centered at a radius of $\sim80$~au
as suggested by \cite{Ricci2015}. The disc extends from nearly $40$ to
180~au, being one of the widest discs resolved at millimetre
wavelengths \citep{Matra2018mmlaw}. We compute the total flux by
integrating the disc emission inside an elliptical mask with a 180~au
semi-major axis and the same aspect ratio as the disc (see
\S\ref{sec:parmodel}). We find a total flux of $16.1\pm1.6$ and
$34.4\pm3.5$~mJy at 1.1 and 0.86~mm (including 10\% absolute flux
uncertainties), leading to a spatially unresolved millimetre spectral
index of $2.64\pm0.48$, consistent with results from
\cite{Ricci2015atca} which combined ALMA and ATCA observations. The
total flux in band 6 is consistent with that measured by
\cite{Ricci2015} at similar wavelengths and using a more compact ALMA
configuration, thus proving that our band 6 observations do not suffer
from flux loss or miss large scale structure.

In order to study in more detail the disc radial structure, the top
panel of Figure~\ref{fig:radial_profile} shows the radial intensity
profile, computed by azimuthally averaging the disc emission over
ellipses as in \cite{Marino2016, Marino2017etacorvi,
  Marino201761vir}. Both in band 6 and 7, we find that the disc
surface brightness peaks near the disc inner edge at $\sim45$~au, from
which it decreases reaching a minimum at 80~au that is deeper in the
band 7 profile likely due to the higher resolution. Beyond this
minimum, the surface brightness increases until 120~au where it peaks
and then decreases steeply with radius. No significant positive
emission is recovered beyond 180 au. Within the gap, we find that the
radial profile is not symmetric with respect to the minimum, with the
outer section (80-100~au) having a steeper slope than the inner part
(60-80 au), a feature that is present both in band 6 and 7 data. This
is also visible in Figure \ref{fig:almab6b7} and is an important
feature that could shed light on the origin of this gap. We also
compute a spectral index map using multi-frequency clean
(\textsc{nterms}=2) and natural weights. The bottom panel of
Figure~\ref{fig:radial_profile} shows the azimuthally averaged radial
profile of the spectral index ($\alphamm$). From 40 to 150 au the disc
has spectral index of roughly $2.7\pm0.1$ (assuming $\alphamm$ is
constant over radii), consistent with the overall spectral index
estimated above. Note that uncertainties on $\alphamm$ do not include
the 10\% absolute flux uncertainties as we are only interested in
relative differences as a function of radius.

In addition to the outer disc, we detect emission from within 30~au
that peaks near the stellar position in the azimuthally averaged
profile. However, it has a much higher level than the photospheric
emission if we extrapolate this from available photometry at
wavelengths shorter than 10$\mu$m using a Rayleigh-Jeans spectral
index ($F_{\nu \star}=32\pm1$ and $18\pm1$~$\mu$Jy at 0.86 and 1.1 mm,
respectively). Moreover, between 20 and 30~au we also find that
$\alphamm$ has a peak of $\sim3.7$, although still within $2\sigma$
from the average spectral index. In \S\ref{sec:parmodel} we recover
this inner component in more detail after subtracting the disc
emission using a parametric model and we find that it is inconsistent
with point source emission, it is significantly offset from the
stellar position, and unlikely to be a background sub-millimetre
galaxy.

The fact that the disc emission is consistent with being axisymmetric
disfavours the scenario in which a planet is scattered out from the
inner regions opening a gap through secular interactions with the disc
\citep[][see their figure 6]{Pearce2015doublering}. In their model,
spiral density features are present in the planetesimal disc for
hundreds of Myr, which should be imprinted in the mm-sized dust
distribution as well, thus being detectable by our observations. In
\S\ref{sec:parmodel} we fit parametric models to the data to study
with more detail the level of axisymmetry of HD~107146's
disc. Moreover, in \S\ref{sec:nbody} we compare our observations with
N-body simulations of a planet on a circular orbit clearing a gap in a
planetesimal disc.

\subsection{CO J=3-2 and HCN J=4-3}
\label{sec:gas}
Despite the increasing number of CO gas detections in nearby debris
discs \citep[e.g.][]{Dent2014, Moor2015gas, Marino2016,
  Marino2017etacorvi, Lieman-Sifry2016, Matra2017fomalhaut, Moor2017},
no CO v=0 J=3-2 emission is detected around HD~107146 in the dirty
continuum-subtracted data cube. This non detection is, however, not
surprising since HD~107146 is significantly older and fainter than the
other systems in which primordial gas has been found (i.e. the hybrid
discs). Furthermore, the sensitivity of these observations is expected
to be insufficient to detect CO gas if being released through
collisions of volatile-rich planetesimals \citep{Kral2017CO}. We
search more carefully for CO emission by applying a matched filter
technique \citep[][]{Matra2015, Marino2016, Marino2017etacorvi,
  Matra2017fomalhaut} in which we integrate the emission over an
elliptic mask with the same orientation as the disc on the sky (see
\S\ref{sec:model}), but in each spatial pixel we integrate only over
those frequencies (i.e. radial velocities) where line emission is
expected taking into account the Doppler shift due to Keplerian
rotation. For this, we assume a stellar mass of 1~$M_{\sun}$,
inclination of $19\degr$, PA of $153\degr$, and deprojected
minimum and maximum radii of 40 and 150~au, respectively. Using this
method we obtain an integrated line flux $3\sigma$ upper limit of 74
mJy~km~s$^{-1}$. Note that we consider the two possible directions of
the rotation, obtaining similar limits. We also search for emission
that could be present at a specific radius by integrating the emission
azimuthally, however no significant emission is found. In
\S\ref{dis:gas}, we use this total flux upper limit to estimate an
upper limit on the CO gas mass that could be present in the disc.


Similarly, we search for HCN emission, finding no significant
emission. Based on this non detection we place a $3\sigma$ upper limit
of 91 mJy~km~s$^{-1}$, which we also use in \S\ref{dis:gas} to
constrain its abundance in planetesimals in this system. HCN is of
particular interest for exocometary studies as, besides being abundant
in Solar System comets \citep{Mumma2011}, it has recently been
suggested that it could play a key role for prebiotic chemistry in
habitable planets \citep{Patel2015, Sutherland2017}.






\section{Modelling}
\label{sec:model}

In this section we model the data using parametric models to constrain
the density distribution of solids in the system
(\S\ref{sec:parmodel}), and N-body simulations of a planet embedded in
a planetesimal disc tailored to HD~107146 to constrain the mass and
orbit of a putative planet carving the observed gap
(\S\ref{sec:nbody}). In both approaches we model the central star as a
G2V type star with a mass of 1~$M_{\sun}$, an effective temperature of
5750~K and a radius of 1~$R_{\sun}$.

\subsection{Parametric model}
\label{sec:parmodel}

We first use a set of parametric models to study the underlying
density distribution of mm-sized dust in the system, which we fit
directly to the observed visibilities as in \cite{Marino2016,
  Marino2017etacorvi, Marino201761vir}. Inspired by the radial profile
of the dust emission (Figure \ref{fig:radial_profile}), we first
choose as a disc model an axisymmetric disc with a surface
density that is parametrized as a triple power law that divides the
disc into an inner edge, an intermediate section (where the bulk of
the dust mass is) and an outer edge. On top of this, the triple power
law surface density distribution has a gap, which we parametrise with
a Gaussian profile, to reproduce the depression seen in the ALMA
data. This parametrization introduces a total of 9 parameters that
define the surface density as follows,
\begin{small}
  \begin{eqnarray}
    &\Sigma(r)=&\Sigma_0 f_\mathrm{gap}(r) \begin{cases}
      \left(\frac{r}{r_{\min}}\right)^{\gamma_1}  & \text{$r<r_{\min}$}, \\
      \left(\frac{r}{r_{\min}}\right)^{\gamma_2}  & \text{$r_{\min}<r<r_{\max}$}, \\
      \left(\frac{r}{r_{\max}}\right)^{\gamma_3}\left(\frac{r_{\max}}{r_{\min}}\right)^{\gamma_2}  & \text{$r>r_{\max}$}, \\
   \end{cases} \label{eq:self-stir}\\
   &f_\mathrm{gap}(r)=& 1-\delta_\mathrm{g}\exp\left[-\frac{(r-r_\mathrm{g})^2}{2\sigma_\mathrm{g}^2}\right],
\end{eqnarray}
\end{small}
where $\rmin$ and $\rmax$ are the inner and outer radii of the disc,
$\gamma_{1,2,3}$ determine how the surface density varies interior to
the disc inner radius, within the disc and beyond the disc outer
radius. The gap is parametrized with a fractional depth $\dgap$, a
center $\rgap$ and a full width half maximum (FWHM)
$\wgap=2\sqrt{2\ln(2)}\siggap$. We leave as a free parameter the total
dust mass $\Md$, which is the surface integral of
$\Sigma(r)$. Although the disc is close to face on, we still model the
dust distribution in three dimensions adding the scale height $h$ as
an extra parameter (i.e. vertical standard deviation of $hr$), and
imposing a prior of $h>0.03$.

We solve for the dust equilibrium temperature and compute images at
0.86 and 1.13 mm using
\textsc{RADMC-3D}\footnote{http://www.ita.uni-heidelberg.de/~dullemond/software/radmc-3d/}. We
assume a weighted mean dust opacity corresponding to dust grains made
of a mix of astrosilicates \citep{Draine2003}, amorphous carbon and
water ice \citep{LiGreenberg1998}, with mass fractions of 70\%, 15\%
and 15\%, respectively, and assuming a size distribution with an
exponent of -3.5 and minimum and maximum sizes of 1$\mu$m and
1cm. This translates to a dust opacity of 1.5~cm$^2$~g$^{-1}$ at 1.1
mm. We note that these choices in dust composition and size
distribution have no significant effect on our modeling apart from the
derived total dust mass. We then use these images to compute model
visibilities at the same \emph{uv} points as the 12m band 6, 12m band
7 and ACA band 7 observations by taking the Fast Fourier Transform
after multiplying the images by the corresponding primary
beam. Additionally, we leave as free parameters the disc inclination
($i$), position angle (PA), RA and Dec offsets for the three
observation sets, and a disc spectral index ($\alphamm$) that sets the
flux at 0.86 mm given the dust mass and opacity at 1.1 mm, i.e. the
size distribution is assumed to be the same throughout the disc. In
total, our model has 19 parameters, 10 for the density distribution
and 9 for the disc centre, orientation, and spectral index.

To find the best fit parameters we sample the parameter space using
the \textsc{PYTHON} module \textsc{emcee}, which implements Goodman \&
Weare's Affine Invariant MCMC Ensemble sampler
\citep{GoodmanWeare2010, emcee}. The posterior probability
distribution is defined as the product of the likelihood function
(proportional to $\exp[-\chi^{2}/2]$) and prior distributions which we
assume uniform, although we impose a lower limit for $h$ of 0.03 due
to model resolution constraints. In computing the $\chi^2$ over the
three visibility sets we applied three constant re-weighting factors
for band~6, 12m band~7 and for ACA band~7 that ensures that the final
reduced $\chi^2$ of each of the three sets is approximately 1 without
affecting the relative weights within each of these data sets that are
provided by ALMA. The re-scaling is necessary as the absolute
uncertainty of ALMA visibilities can be offset by a factor of a few,
even after re-weighting the visibilities with the task \textsc{statwt}
in CASA 4.7. These factors could be alternatively left as free
parameters by adding an extra term to the likelihood function, however
we find no differences in our results compared to leaving them fixed
during multiple tries. Therefore we opt for leaving them fixed.

\begin{table}
  \centering
  \caption{Best fit parameters of the ALMA data for the different
    parametric models. The quoted values correspond to the median,
    with uncertainties based on the 16th and 84th percentiles of the
    marginalised distributions or upper limits based on 95th
    percentile.}
  \label{tab:mcmc}
  \begin{tabular}{lrl} 
    \hline
    \hline
    Parameter & best fit value & description\\
    \hline
    \multicolumn{3}{l}{3-power law + Gaussian gap}\\
    \hline
    $\Md$ [$M_{\earth}$]& $0.250\pm0.004$ & total dust mass \\
    $\rmin$ [au] & $46.6^{+1.4}_{-1.5}$ & disc inner radius \\
    $\rmax$ [au] & $135.6^{+1.1}_{-1.2}$ & disc outer radius \\
    $\gamma_1$   & $2.6^{+0.3}_{-0.2}$ & inner edge's slope\\
    $\gamma_2$   & $0.26^{+0.08}_{-0.10}$ & disc slope\\
    $\gamma_3$   & $-10.5^{+0.9}_{-1.0}$  & outer edge's slope \\
    $\rgap$ [au] & $75.5^{+1.1}_{-1.2}$ & radius of the gap \\
    $\wgap$ [au] & $38.6^{+4.5}_{-3.6}$& FWHM of the gap\\
    $\dgap$      & $0.52^{+0.03}_{-0.02}$ & fractional depth of the gap\\
    $h$          & $0.12^{+0.04}_{-0.05}$ & scale height\\
    PA [$\degr$] & $153\pm3$& disc PA\\
    $i$ [$\degr$]& $19.3\pm1.0$ & disc inclination from face-on\\
    $\alphamm$   & $2.57\pm0.11$& millimetre spectral index \\
    \hline
    \multicolumn{3}{l}{Step gap} \\
    \hline
    $\rgap$ [au] & $75.4^{+0.8}_{-0.7}$ & radius of the gap\\
    $\wgap$ [au] & $42.2^{+1.7}_{-2.2}$ & width of the gap \\
    $\dgap$  &   $0.43\pm0.02$  & depth of the gap \\
    \hline
    \multicolumn{3}{l}{Eccentric disc} \\
    \hline
    $\ed$ &  <0.03 &  disc global eccentricity\\
    \hline
    \multicolumn{3}{l}{Inner component} \\
    \hline
    $\rmin$ [au] & $41.9^{+1.2}_{-1.4}$ & disc inner radius \\
    $\gamma_1$   & $11.6^{+3.0}_{-2.7}$ & inner edge's slope\\
    $\gamma_2$   & $0.03^{+0.19}_{-0.26}$ & disc slope\\
    $\rgap$ [au] & $72.1^{+2.2}_{-2.9}$ & radius of the gap \\
    $\wgap$ [au] & $51^{+12}_{-8}$& FWHM of the gap\\
    $\dgap$      & $0.58^{+0.08}_{-0.06}$ & fractional depth of the gap\\
    $\Mc$ [$M_{\earth}$] & $3.0^{+0.9}_{-0.6}\times10^{-3}$  & dust mass inner component\\
    $\rc$  [au] & $19.3^{+2.8}_{-2.8}$ & radius of inner component \\
    $\deltarc$ [au] & $35.8^{+9.1}_{-6.7}$ & radial width of inner component \\
    $\omegac$ [$\degr$] &  $85^{+9}_{-9}$ & PA of inner component south of disc PA,\\
    & & and in the disc plane  \\
    $\sigphi$ [$\degr$] & $94^{+15}_{-12}$ & azimuthal width of inner component \\
    \hline
    
  \end{tabular}
 
\end{table}

In Table \ref{tab:mcmc} we present the best fit parameters of our
3-power law model with a Gaussian gap. We find a disc inner radius of
47~au that is significantly larger than the inner edge of
$\sim25-30$~au derived by \cite{Ricci2015}. This is because in their
model they considered a sharp inner edge, while in ours we allow for
the presence of dust within this inner radius, but decreasing towards
smaller radii. We find that $r_{\min}$ matches well the radius at
which the surface density peaks as seen in Figure
\ref{fig:radial_profile}. Similarly, our estimate of the outer radius
(136 au) is significantly smaller than their outer edge (150~au). We
find that the disc inner edge has a power law index (slope hereafter)
of $\sim2.6$, while the outer edge is much steeper with a slope of
$-11$. The intermediate component has a very flat slope of $\sim0.3$,
$2.7\sigma$ flatter than the previous estimate (0.59), although the
difference could be simply due to different parametrizations,
i.e. considering a three vs a single power law parametrization or
leaving the gap's depth as a free parameter vs fixed to 1. Note that
the slope derived from the intermediate component does not imply that
the mass surface density of planetesimals is flat. In fact, from
collisional evolution models we expect the surface density of mm-sized
dust and optical depth to have a lower slope than the total mass
surface density in regions of the disc where the largest planetesimals
are not yet in collisional equilibrium \citep[i.e. have a lifetime
  longer than the age of the system,][]{Schuppler2016,
  Marino201761vir, Geiler2017}. Moreover, we also expect that in the
inner regions where the largest planetesimals are in collisional
equilibrium, the surface density of material should have a slope close
to $7/3$ \citep{Wyatt2007hotdust, Kennedy2010}, as we find interior to
$r_{\min}$. This suggests that the regions interior to 47~au might be
relatively depleted of solids simply due to collisional evolution
rather than clearing by planets or inefficient planetesimal
formation. We compare the derived inner radius and surface density of
millimetre grains ($\sim3\times10^{-6}$~\Me~au$^{-2}$ at 50~au) with
collisional evolution models by \cite{Marino201761vir}, which include
how the size distribution evolves at different radii, to estimate the
maximum planetesimal size and initial total surface density of
solids. We find that the best match has a maximum planetesimal size of
$\sim10$~km and an initial disc surface density of
0.015~\Me\ au$^{-2}$ at 50~au, i.e. 5 times the surface density of the
Minimum Mass Solar Nebula \citep{Weidenschilling1977mmsn, Hayashi1981}
extrapolated to large radii, or a total solid disc mass of
300~\Me. This implies a very massive initial disc and efficient
planetesimal formation at large radii. These conclusions assume,
however, that the observed structure in the surface brightness profile
arise from collisional evolution neglecting alternative
origins. Finally, regarding the disc orientation, we find values that
are consistent with previous estimates, but with tighter constraints
(see Table \ref{tab:mcmc}).

Our results show that the gap is centered at $75.5^{+1.1}_{-1.2}$~au,
consistent with the previous estimate. However, we find a FWHM of
$\sim$40~au that is much larger than the previous estimate of 9~au,
likely due to \cite{Ricci2015} assuming a gap depth of 100\%. Instead,
we fit the depth of the gap finding a best fit value of 0.5. We also
fit an alternative model in which the gap is a step function with a
constant depth, finding best fit values that are similar to the model
with the Gaussian gap (see Table \ref{tab:mcmc}). The difference in
widths between the previous study and this work is interesting as
\cite{Ricci2015} derived the mass of a putative planet clearing a gap
in the disc based on the gap's width and assuming that it should be
roughly equal to the planet's chaotic zone. A 3-4 times wider gap
would imply a much higher planet mass as the width of the chaotic zone
scales as $\Mp^{2/7}$ \citep{Wisdom1980, Duncan1989}. Such planet mass
estimates assume that the system is in steady state, that the gap is
devoid of material, and that the gap's width is simply equal to the
chaotic zone's width. However, given the age of the system
($\sim$80-200~Myr), and that any planet may be younger, it is
reasonable to consider that the distribution of particles in the
planet's vicinity could still be evolving. Therefore, instead of using
the gap's width to estimate a planet mass, in \S\ref{sec:nbody} we
estimate this by comparing with N-body simulations tailored to
HD~107146.

\begin{figure}
  \centering
  \includegraphics[trim=0.0cm 0.0cm 0.0cm 0.0cm, clip=true,
    width=1.0\columnwidth]{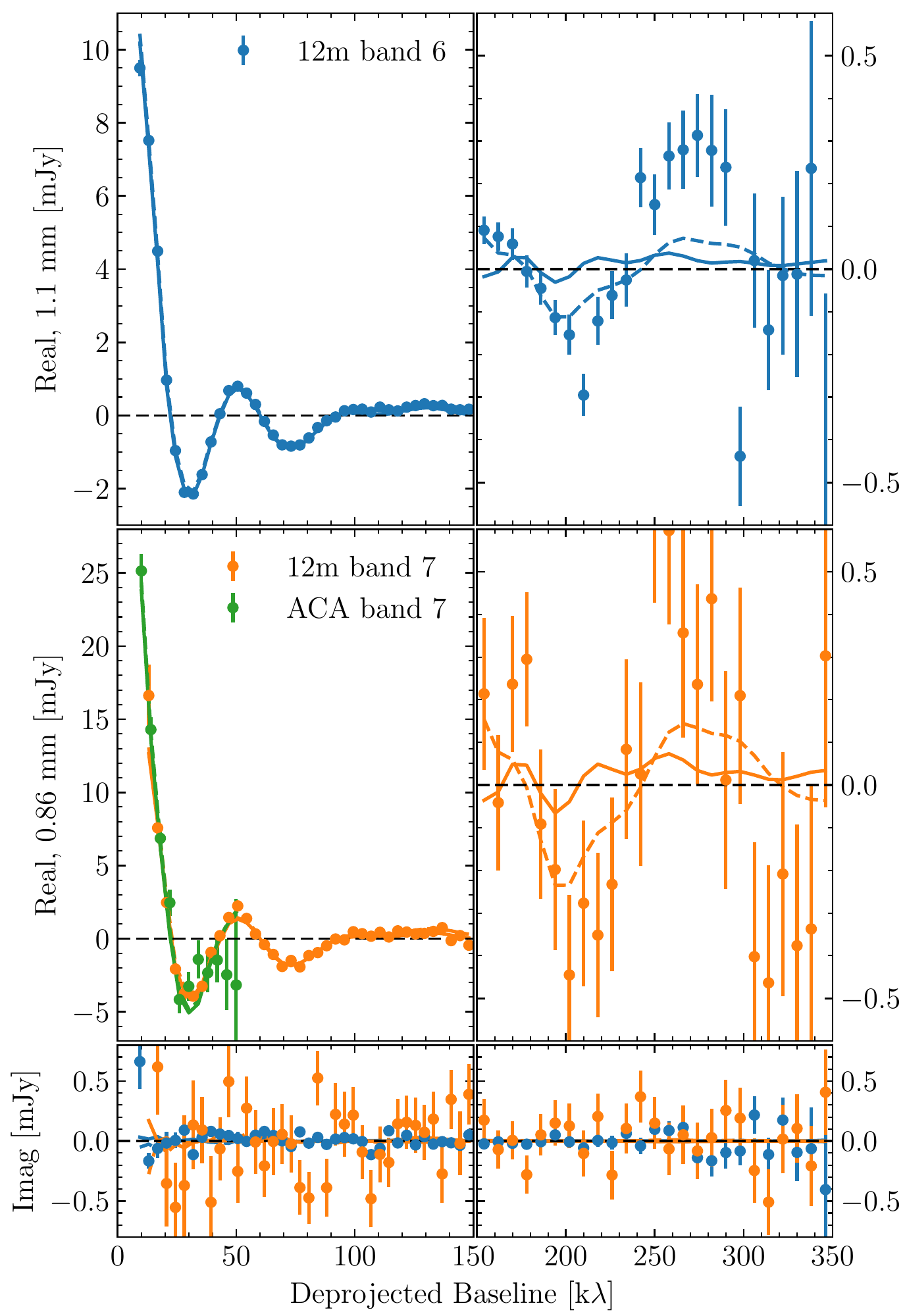}
 \caption{Deprojected and binned visibilities assuming a disc position
   angle of $153\degr$ and inclination of 19$\degr$. The real
   components of the band 6 and band 7 data are presented in the top
   and middle panels, respectively. The imaginary components of the
   12m data are displayed in the lower panels.  The errorbars
   represent the binned data with their uncertainty estimated as the
   standard deviation in each bin divided by the square root of the
   number of independent points. The continuous line shows the triple
   power law best fit model with a Gaussian gap. The dashed line
   represents the best fit model with the same parametrization, but
   with an additional inner component. Note that the scale in the left
   and right panels is different.}
 \label{fig:vis}
\end{figure}

\begin{figure}
  \centering
 \includegraphics[trim=1.5cm 0.0cm 0.5cm 0.0cm, clip=true, width=1.0\columnwidth]{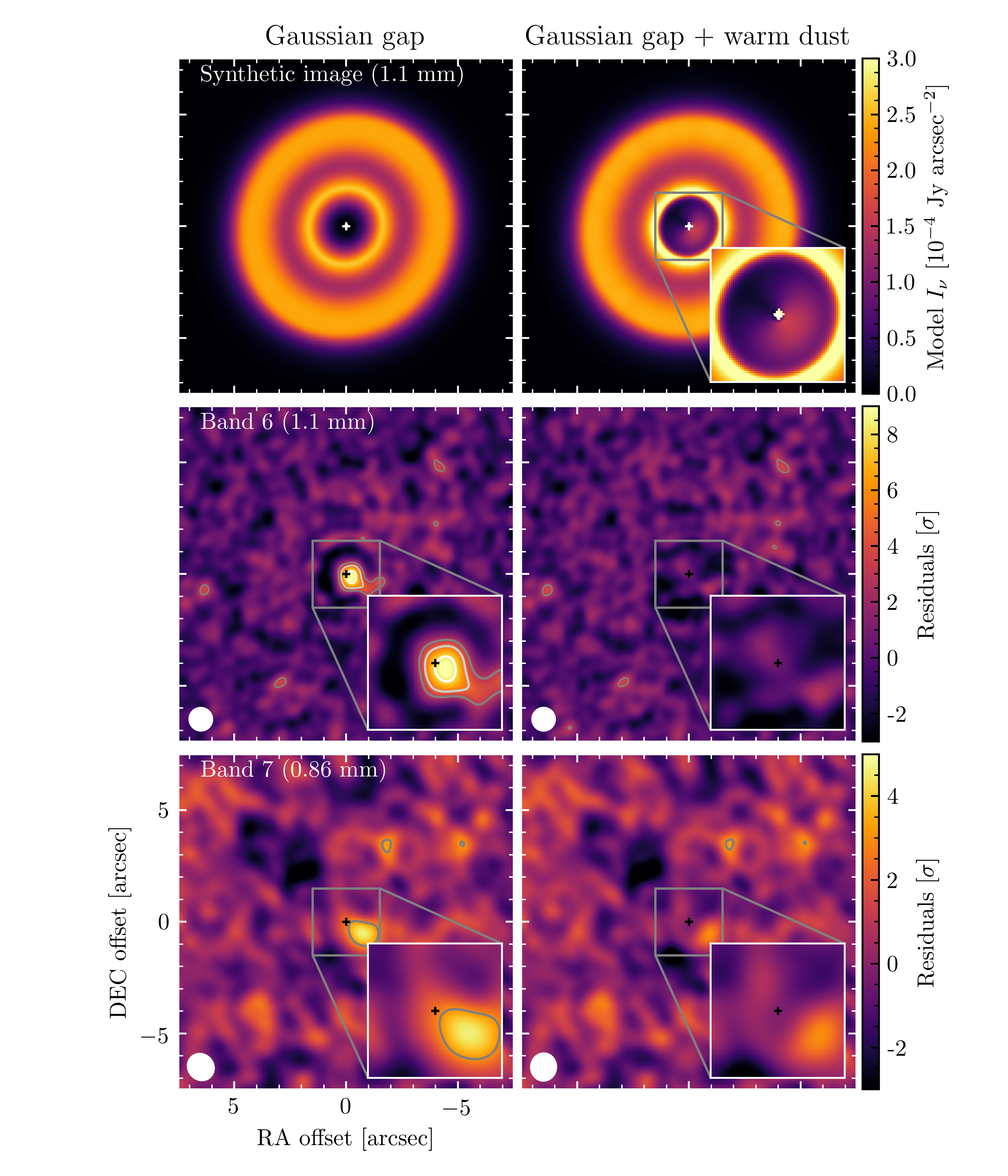}
 \caption{Best fit model images (1.1 mm) and dirty map of
   residuals. \emph{\textbf{Left column:}} 3-power law model with a
   Gaussian gap. \emph{\textbf{Right column:}} 3-power law model with
   a Gaussian gap and a warm inner component. The residuals of band 6
   (middle row) and band 7 (12m+ACA, bottom row) are computed using
   natural weights and with an outer taper of $0\farcs8$, leading to a
   synthesised beam of $0\farcs80\times0\farcs79$ and
   $1\farcs0\times0\farcs9$ and an rms of 6.1 and 41
   $\mu$Jy~beam$^{-1}$, respectively. Contours represent 3, 5 and
   8$\sigma$.}
 \label{fig:res}
\end{figure}

Using the best fit PA and $i$ we deproject the observed visibilities
and bin them to compare them with our axisymmetric model in Figure
\ref{fig:vis} (continuous line). The 12m and ACA band 7 observations
are consistent with each other within errors, with some systematic
differences due to the different primary beams. We find that our best
fit model fits well both the real and imaginary components, with the
imaginary part being consistent with zero as expected for an
axisymmetric disc. However, we find a significant deviation between
the real components of the 12m band 7 data and model at around
50~k$\lambda$ (corresponding to an angular scale of $4\arcsec$), with
the model visibility predicting a slightly lower value than the
data. This could be due to large scale variations in the spectral
index as the same model fits well those baselines in the band 6
data. This deviation is the same when comparing the model with a step
gap, which has an indistinguishable visibility profile for baselines
shorter than 200~k$\lambda$. Beyond 150~k$\lambda$ we also find
deviations between the data and model, expected since as shown in
Figure \ref{fig:radial_profile} the radial profile seems to have
structure that is more complex than a simple Gaussian gap.

A more intuitive way to study the goodness of fit of our model is to
look at the dirty map of the residuals. The left column of Figure
\ref{fig:res} shows the dirty maps of the residuals after subtracting
our best fit model from the data (in visibility space). These are
computed using natural weights, plus an outer taper of $0\farcs8$ for
band 7 to obtain a synthesised beam of similar size compared to the
band 6 beam. Both band 6 and 7 residual images show significant
residuals (with peaks of 9 and 5 times the rms) within the disc inner
edge and with a peak that is significantly offset from the stellar
position by $\sim0\farcs5$, corresponding to a projected distance of
$\sim15$~au. Moreover, the emission is marginally resolved, extending
radially by more than a beam (i.e. larger than 10~au) and with an
integrated flux significantly higher than the predicted photospheric
emission at these wavelengths assuming a Rayleigh-Jeans spectral index
(18 and 32 $\mu$Jy at 1.1 and 0.86 mm). We estimate the flux of these
components by integrating the dirty maps over a circular region of
$2\arcsec$ diameter, finding an integrated flux of $85\pm13$~$\mu$Jy
and $290\pm70$~$\mu$Jy at 1.1 and 0.86 mm, respectively (without
including 10\% absolute flux uncertainties). Because this inner
emission is resolved, it cannot arise from a compact source such as a
planet or circumplanetary material.

We identify two plausible origins for this inner component. It could
be the same warm dust that was inferred based on Spitzer data
\citep{Morales2011}, as the clump is at a radial distance that is
consistent with the inferred black body radius (5-15~au) from its
spectral energy distribution (SED), although our detection would imply
that the warm component has an asymmetric distribution (see discussion
in \S\ref{dis:warmdust}). On the other hand, the residual could be
also due to a background sub-millimeter galaxy that are often detected
incidentally as part of ALMA deep observations
\citep[see][]{Simpson2015, Carniani2015, Marino201761vir,
  Su2017}. Sub-millimeter galaxies have typical sizes of the order of
1\arcsec\ and spectral indices ranging between $\sim3-5$ at
mm-wavelengths, thus consistent with the observed clump. We estimate a
probability of $\sim50\%$ of finding a sub-millimeter galaxy as bright
as 0.3 mJy at 0.86 mm or 0.1~mJy at 1.1~mm, respectively, within the
entire band 6 and 7 primary beams \citep{Simpson2015,
  Carniani2015}. Hence, the detection of such a bright background
object is not a rare event. However, we find that the probability of
finding such a bright sub-millimeter galaxy at 0.86~mm and 1.1~mm and
co-located with warm dust (i.e within $1\arcsec$ from the star) is
only 0.6\% and 1\%, respectively; therefore favouring the warm dust
scenario. In fact, we do not detect any other compact emission above
$5\sigma$ within the band 7 and 6 primary beams, consistent with the
number counts of sub-millimeter galaxies. We also discard that this
emission could originate from the galaxy detected using HST in 2004,
2005 and 2011 \citep{Ardila2004, Ertel2011, Schneider2014}, as its
position would only have changed by $1.4\arcsec$, and therefore still
lie at $\sim5\arcsec$ ($\sim140$~au) from the star in 2017, and is not
detected in our observations.





\subsubsection{Disc global eccentricity}

As shown by \cite{Pearce2014, Pearce2015doublering}, a planet on an
eccentric orbit can force an eccentricity in a disc of planetesimals
through secular interactions. Here we aim to assess if HD~107146's
debris disc could have a global eccentricity and pericentre, using the
same parametrization as in \cite{Marino2017etacorvi}, i.e. taking into
account the expected apocentre glow \citep{Pan2016}. We find that the
disc is consistent with being axisymmetric, with a 2$\sigma$ upper
limit of 0.03 for the forced eccentricity. We find though that the
marginalised distribution of $\ed$ peaks at 0.02 with a pericentre
that is opposite to the residual inner clump. This peak is likely
produced as the residuals are lower when the disc is eccentric and
with an apocentre oriented towards the clump's position angle due to
apocentre glow. We therefore conclude that the fit is biased by the
inner clump and that the disc is probably not truly eccentric.

\subsubsection{Inner component}
\label{sec:modelin}
In order to constrain the geometry or distribution of the inner
emission found in the residuals, we add an extra inner component by
introducing five additional parameters to our reference parametric
model (3-power law surface density with a Gaussian gap). We
parametrise its surface density as a 2D Gaussian in polar coordinates,
with a total dust mass $\Mc$ and centered at a radius $\rc$ and
azimuthal angle $\omegac$ (measured in the plane of the disc from the
disc PA and increasing in an anti-clockwise direction). The width of this
Gaussian is parametrized with a radial FWHM $\deltarc$ and an
azimuthal standard deviation $\sigphi$. Best fit values are presented
in Table \ref{tab:mcmc}. We find that this inner component is extended
both radially and azimuthally, but concentrated around $19\pm3$~au
and orthogonal to the disc PA ($\omegac\sim90\degr$), as we found in
the residuals of our axisymmetric model. In the right panel of Figure
\ref{fig:res} we present the model image of the best fit model and its
residuals, which are below $3\sigma$ in both band 6 and 7 within the
disc inner edge. We also find that the total dust mass of this inner
component is $3.0^{+0.9}_{-0.6}\times10^{-3}$~\Me\ ($\sim$1\% of the
outer disc mass).

When adding this extra inner component we find that the slope of the
inner edge is steeper and the inner radius smaller compared with our
previous model (symmetric disc model hereafter), as it was probably
compensating for the emission within 30~au with a less steep inner
edge. We also find a slightly smaller gap radius of 72~au, a larger
and deeper gap (51~au wide and 0.58 deep), and a flatter surface
density slope of $0.0\pm0.2$. These differences in the gap's structure
and disc slope are overall consistent within $3\sigma$ with our
previous estimates, but significantly improve the fit at large
baselines as Figure~\ref{fig:vis} shows (dashed blue and orange
lines). These improvements in the fit at large baselines are not due
to the addition of the clump, but due to the different best fit
surface density profile of the outer disc. In fact, the visibilities
of the inner component are negligible beyond 200~k$\lambda$. The rest
of the parameters are consistent within $1\sigma$ with the values
presented for the symmetric disc model. To have a better estimate of
the flux of this inner component, we subtract the new best fit of the
outer component (its inner edge is steeper), finding an integrated
flux over a circular region of $3\arcsec$ diameter of $0.82\pm0.14$
and $0.31\pm0.04$ mJy at 0.86 and 1.1 mm, respectively (including 10\%
absolute flux uncertainties)\footnote{We also measured the integrated
  flux of the inner component using the task \textsc{uvmodelfit} in
  CASA 4.7, obtaining values of $0.35\pm0.03$ and $0.81\pm0.14$~mJy at
  1.1 and 0.86~mm, respectively.}. Note that this flux is a factor two
higher than that estimated from the residuals of the symmetric disc
model, as without the inner component the model tries to compensate
for the emission interior to 40~au. From these fluxes we estimate a
spectral index of $3.3\pm0.6$, thus still consistent with the typical
observed spectral indices of debris discs. Based on this new flux
estimate at 0.86~mm, we find an even lower probability of 0.1\% and
0.3\% of finding a sub-mm galaxy as bright as this inner emission at
0.86 and 1.1~mm, respectively, and within $1\arcsec$ from the star
\citep{Simpson2015, Carniani2015}. These results also confirm that the
peak of this inner component is significantly offset from the star and
it is incompatible with an axisymmetric inner component. If this
emission is produced by warm dust, then it could bring valuable
insights into the origin of warm dust emission in general, as it is
inconsistent with an axisymmetric asteroid belt (see
\S\ref{dis:warmdust}).

Because the new estimate of the inner edge slope is too steep to be
consistent with being set by collisional evolution, the maximum
planetesimal size is only constrained to be $\gtrsim10$~km. Despite
this, the relative brightness between the inner and outer components
can still be explained simply by collisional evolution. This has also
been found for other systems with warm dust components,
e.g. q$^1$~Eridani \citep{Schuppler2016}, suggesting the presence of
planets clearing the material in between \citep{Shannon2016}.

\clearpage

\begin{figure*}
  \centering
 \includegraphics[trim=0.3cm 0.3cm 0.3cm 0.3cm, clip=true, width=0.8\textwidth]{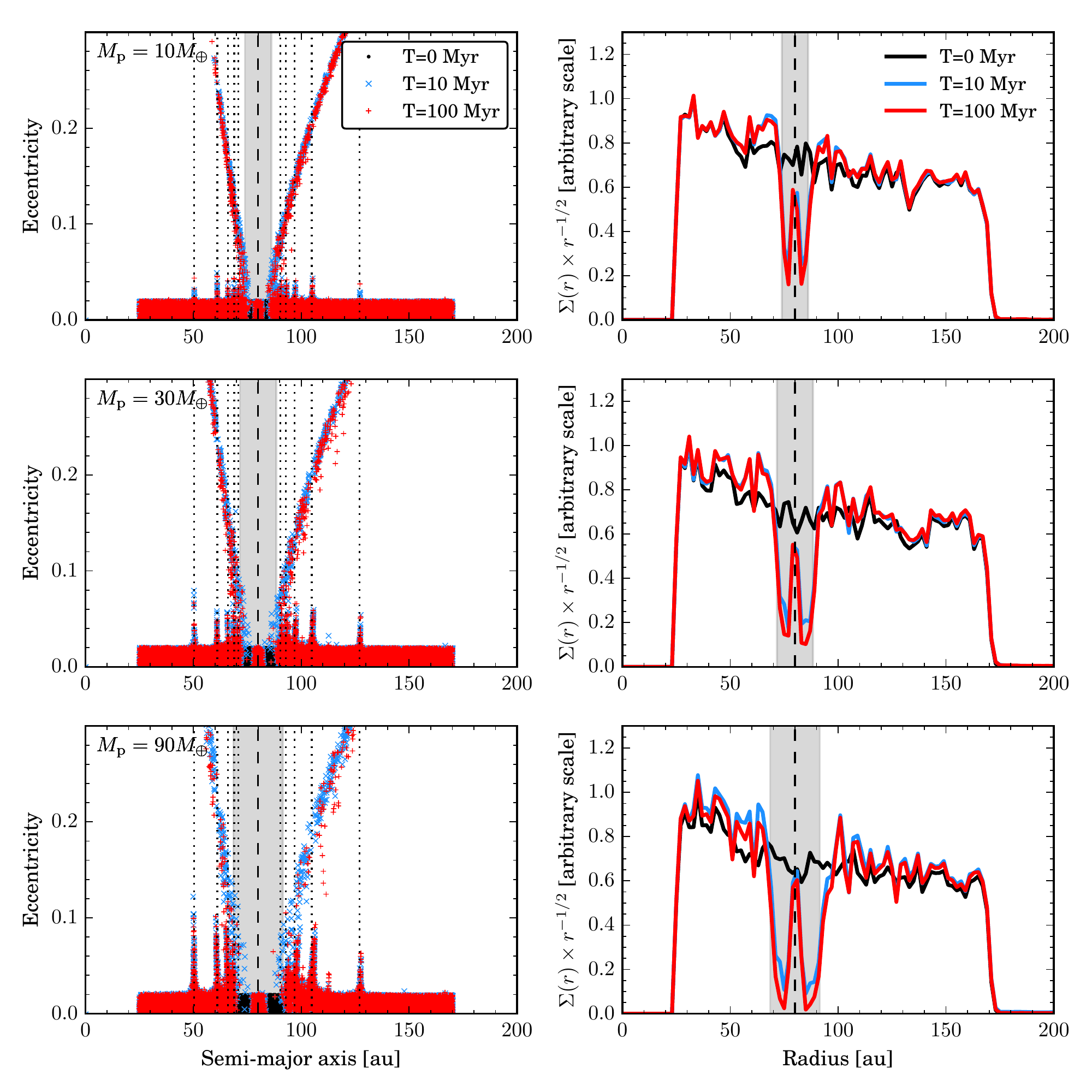}
 \caption{Evolution of massless particles in a system with a single
   planet on a circular orbit at 80 au. \emph{\textbf{Left column:}}
   eccentricity and semi-major axis of particles after 0, 10 and 100
   Myr of evolution. The dotted and dashed vertical lines represent
   first order mean motion resonances and the semi-major axis of the
   planet. \emph{\textbf{Right column:}} Surface brightness of
   particles assuming an initial surface density proportional to
   $r^{1/4}$ and a dust temperature profile decreasing with radius as
   $r^{-1/2}$. The grey shaded region represents the chaotic zone
   approximated by $\ap\pm1.5\ap(\Mp/M_{\star})^{2/7}$. The top, middle and
   bottom panels show systems with planet masses of 10, 30 and 90~\Me,
   respectively.}
 \label{fig:nbody}
\end{figure*}

\subsection{N-body simulations}
\label{sec:nbody}

In this section we compare the observations with dynamical N-body
simulations of a planet embedded in a planetesimal disc.  To simulate
the gravitational interactions between the planet and particles we use
the N-body software package \textsc{REBOUND} \citep{rebound}, using
the hybrid integrator \textsc{Mercurius}\footnote{We also used Hermes
  in several trial runs; however, we decided to use Mercurius instead
  because with Hermes we obtained results that differ significantly
  from other integrators such as ias15, whfast and the hybrid
  integrator within Mercury. We found that some particles outside the
  chaotic zone were driven to unstable orbits and did not conserve
  their Tisserand parameter. The possibility of a potential bug in
  Hermes was confirmed by private communication with Hanno Rein.} that
switches from a fixed to a variable time-step when a particle is
within a given distance from the planet (here chosen to be 8 Hill
radii). The fixed time step is chosen to be 4\% of the planet's
period, which is lower than 17\% of the orbital period of all
particles in the simulation. Although the total mass of the disc could
be tens of \Me, and thus comparable with the mass of the simulated
planets, we assume particles have zero or a negligible mass. In
\S\ref{sec:massparticles} we discuss the effect of considering a
massive planetesimal disc on planet disc interactions.


Particles are initially randomly distributed in the system with a
uniform distribution in semi-major axis ($a$) between 20 and 170~au
(i.e. with a surface density proportional to $r^{-1}$), with an
eccentricity and inclination uniformly distributed between 0-0.02 and
0-0.01 radians, respectively. We use a total number of $10^{4}$
particles, sufficient to sample the 150~au span in semi-major axes and
recover smooth images of the disc density distribution. The planet is
placed on a circular orbit at 80~au and assumed to have a bulk density
of 1.64 g~cm$^{-3}$ (Neptune's density). We integrate the evolution of
the system up to 200~Myr, roughly the upper limit for the age of the
system. We run a set of simulations with planet masses varying from 10
to 100~$M_{\earth}$ with a 5~$M_{\earth}$ spacing.

In order to translate the outcome of these simulations to density
distributions used to produce synthetic images, we populate the orbits
of each particle with 200 points randomly distributed uniformly in
mean anomaly as in \cite{Pearce2014}, but in the frame co-rotating
with the planet in order to see if there are resonant structures
(e.g. loops or Trojan regions). We find however no significant
azimuthal structures due to first or second order resonances in the
derived surface density of particles. We weight the mass of each
particle based on its initial semi-major axis to impose an initial
surface density proportional to $r^{\gamma}$ between a minimum and
maximum semi-major axis ($a_{\min}, a_{\max}$). Finally, because we
only run simulations with a fixed planet semi-major axis of 80~au to
save computational time, we scale all the distances in the output of
the simulation and leave $\ap$ as a free parameter (only varying
roughly between 75-80 au). This linear scaling is motivated by the
fact that some of the features that we are interested in should scale
with semi-major axis (e.g. Hill radius, mean-motion resonances,
chaotic zone), although some other important quantities, such as the
scattering diffusion timescale \citep{Tremaine1993}, have a dependency
on $a$ that is different from linear. In Appendix \ref{appendix} we
test the validity of the scaling approximation for the narrow range of
planet semi-major axis that we explore.


In Figure \ref{fig:nbody} we present, as an example, the evolution of
$a$, $e$ and the surface density of particles for planet masses of 10,
30 and 90~\Me. The width of the chaotic zone is overlayed in grey
(note that this is a region in semi-major axis rather than radius), to
compare with the gap in the surface density cleared by the
planet. Although by 100~Myr the chaotic zone is almost empty of
material (except for particles in the co-rotation zone) the surface
density is not zero inside the gap as some particles have apocentres
or pericentres within this region while being scattered by the
planet. Some particles also remain on stable tadpole or Trojan orbits
until the end of the simulation, creating an overdensity within the
gap at 80 au. Interior and exterior to the planet, a small fraction of
particles are in mean motion resonance with the planet and have their
eccentricities increased. In agreement with previous work, we find
that the gap's width approximates to the chaotic zone, and its width
does not vary after 10 Myr for the planet masses explored.

\begin{figure}
  \centering
 \includegraphics[trim=0.0cm 0.0cm 0.0cm 0.0cm, clip=true, width=1.0\columnwidth]{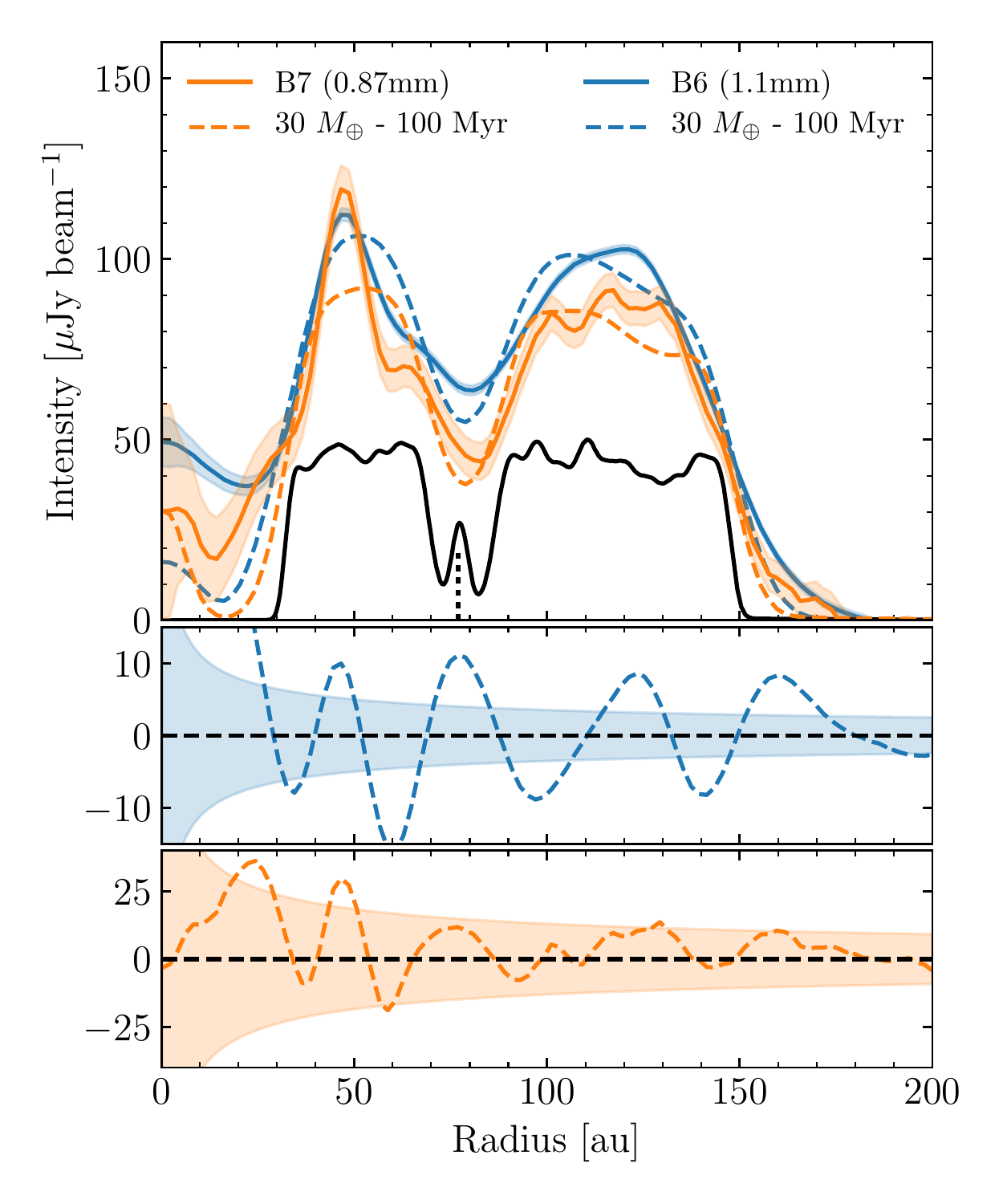}
 \caption{\textbf{\emph{Top:}} Average intensity profile computed
   azimuthally averaging the disc emission over ellipses oriented as
   the disc in the sky. The blue and orange lines are obtain from the
   band 6 and band 7 Clean images (continuous lines) and simulated
   Clean images based on an N-body simulation of one 30~\Me planet
   (dashed lines), using Briggs (robust=0.5) weights. The black
   continuous line represents the surface brightness profile with a
   1~au resolution and displayed at an arbitrary scale. The vertical dotted line
   represents the orbital radius of the
   planet. \textbf{\emph{Middle:}} Azimuthally averaged residuals in
   band 6. \textbf{\emph{Bottom:}} Azimuthally averaged residuals in
   band 7. The shaded areas represent the 68\% confidence region in
   the top panel and 99.7\% confidence region in the middle and bottom
   panel, over a resolution element (18 au for band 6 and 13 au for
   band 7).}
 \label{fig:radial_profile_nbody}
\end{figure}

To compare with observations, we use snapshots of the simulations at
50, 100 and 200 Myr since the age of the system is uncertain and could
vary roughly within this range. These ages assume that the putative
planet formed (or grew to its current mass) early during the evolution
of this system, rather than recently, or that this is the time since
the planet formed. For each assumed epoch, we explore the parameter
space using the same MCMC technique as in \S\ref{sec:parmodel},
varying $\amin$ and $\amax$, the surface density exponent $\gamma$,
the semi-major axis of the planet, and its mass by interpolating the
resulting surface density between two neighbouring simulations. We
also leave as free parameters the disc orientation, pointing offsets
and spectral index, and instead of varying the disc scale height we
assume a flat disc. We find a best fit mass of $30\pm5$~\Me\ and
semi-major axis of $77\pm1$~au, independently of the assumed age since
after 10~Myr there is no significant evolution in the orbits of
particles near the planet (see Figure \ref{fig:nbody}). We find,
however, that our best fit cannot reproduce well the width and depth
of the gap. This is illustrated in the top and middle panels in Figure
\ref{fig:radial_profile_nbody}, where the model has a narrower gap
compared with the band 6 radial profile, and thus overall larger
residuals. Although larger planet masses could produce wider gaps,
these would also be significantly deeper and thus inconsistent with
the observations. Moreover, the planet gap could be even deeper if,
for example, we assume a different starting condition with a depleted
surface density within the planet's feeding zone as it accreted a
large fraction of that material while growing. To test this, we repeat
the fitting procedure, but removing those particles that start the
simulation within two Hill radii from the planet's orbit (planet's
feeding zone). We find a slightly lower best fit planet mass of
$26\pm3$~\Me, but overall the fit is worse with a difference of
$\sim300$ in the total $\chi^2$, strongly preferring the model with
particles starting near the planet. Therefore, we conclude that a
single planet on a circular orbit that was born within the outer disc
is unable to explain the ALMA observations.

\section{Discussion}
\label{sec:dis}

\subsection{The gap's origin}

As discussed in \S\ref{sec:intro}, we identify multiple scenarios
where a single planet might open a gap in a planetesimal disc. Below
we discuss these and how well they could fit the data.

\subsubsection{Single planet on a circular orbit}
As we showed in \S\ref{sec:nbody}, a single planet on a circular orbit
and formed inside the gap could clear its orbit of debris through
scattering, opening a gap with a width similar to the chaotic
zone. Such a planet, however, needs to be very massive to produce a
$\sim40$~au wide gap ($\gtrsim3$~\Mjup), which, on the other hand,
results in a gap that is significantly deeper than our observations
suggest. Based on these constraints we find that a 30~\Me\ planet is
the best trade-of between the gap's width and depth.

If, however, the planet could migrate (inwards or outwards) then these
two observables could become compatible. A low mass planet that
migrated through the disc could carve a sufficiently wide gap to
explain the $\sim40$~au gap's width. Migration could also explain the
relative depth of $\sim50\%$ that is observed as after migrating, the
planet would also leave behind debris that was scattered onto excited
orbits, but that do no longer cross the planet's orbit
\citep[e.g.][]{Kirsh2009}, therefore the gap would still contain a
significant fraction of the original material. Planet migration could
be induced by planetesimal scattering, which is discussed in
\S\ref{sec:massparticles}.

\subsubsection{Multiple planets on circular orbits}

\begin{figure}
  \centering
 \includegraphics[trim=0.0cm 0.0cm 0.0cm 0.0cm, clip=true, width=1.0\columnwidth]{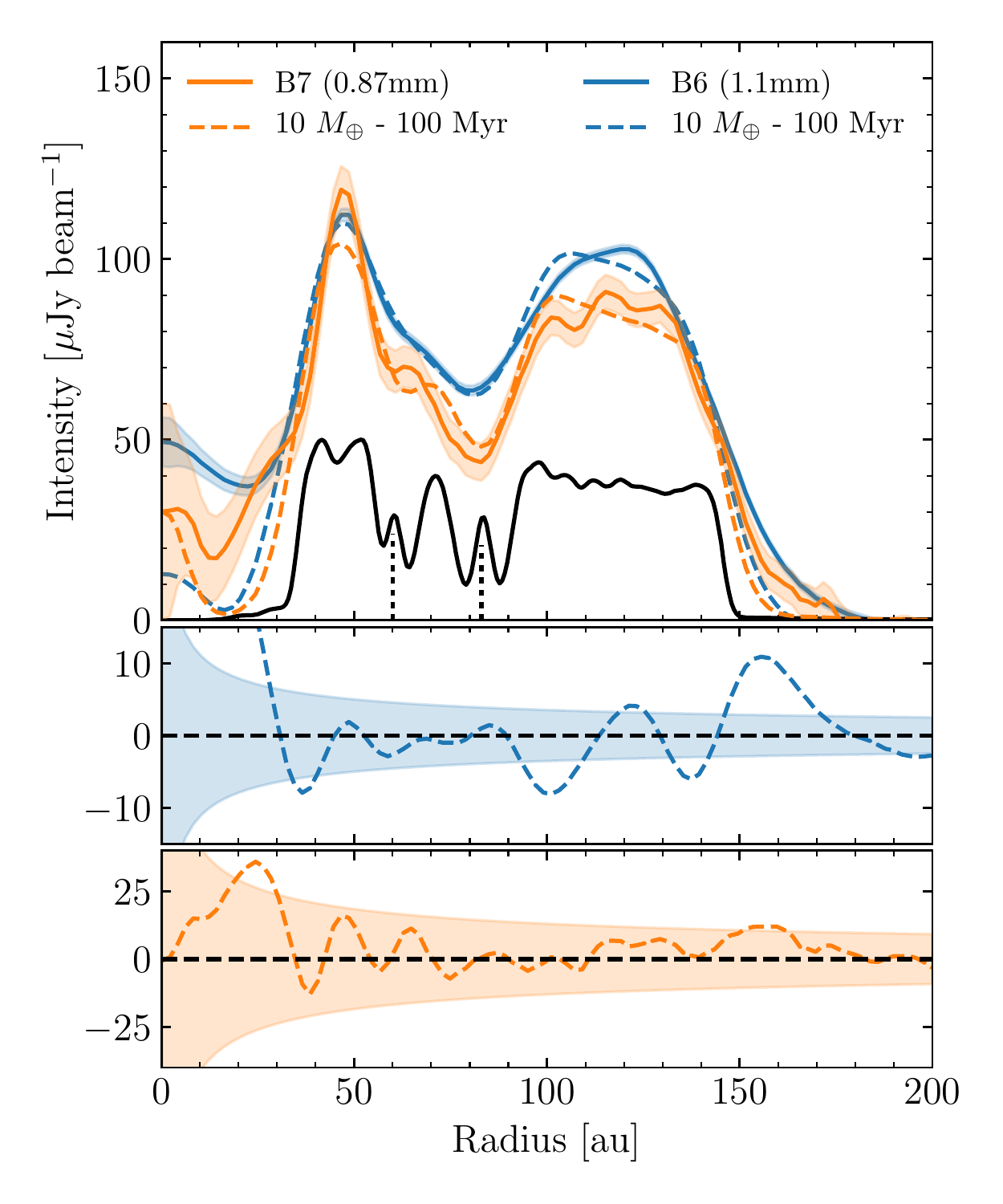}
 \caption{\textbf{\emph{Top:}} Average intensity profile computed
   azimuthally averaging the disc emission over ellipses oriented as
   the disc in the sky. The blue and orange lines are obtain from the
   band 6 and band 7 Clean images (continuous lines) and simulated
   Clean images based on an N-body simulation of two 10~\Me planets
   (dashed lines), using Briggs (robust=0.5) weights. The black
   continuous line represents the surface brightness profile with a
   1~au resolution and displayed at an arbitrary scale. The vertical
   dotted line represents the orbital radii of the
   planets. \textbf{\emph{Middle:}} Azimuthally averaged residuals in
   band 6. \textbf{\emph{Bottom:}} Azimuthally averaged residuals in
   band 7. The shaded areas represent the 68\% confidence region in
   the top panel and 99.7\% confidence region in the middle and bottom
   panel, over a resolution element (18 au for band 6 and 13 au for
   band 7).}
 \label{fig:radial_profile_nbody_2plt}
\end{figure}

If we assume planet migration is negligible (e.g. disc mass around the
planet's orbit is much lower than its mass), then multiple planets
would need to be present to carve such a wide and shallow gap. Let us
assume that the gap was carved by multiple equal mass planets orbiting
between 60 and 90~au. As shown by \cite{Shannon2016} and in
\S\ref{sec:nbody}, the width of the gap and age of the system place
tight constraints on the minimum mass of a planet to clear the region
surrounding its orbit, and on the maximum number of planets that could
be orbiting within the gap based on a stability criterion. Given
HD~107146's age limits of $\sim50$ and 200~Myr, only planets with
masses greater than 10~\Me\ would have enough time to clear their
orbits via scattering. On the other hand, only planet masses
$\lesssim30$~\Me\ can create a gap that is not too deep compared with
our observations. Given this range of planet masses, we estimate that
a maximum of three 10~\Me\ planets could orbit between 60 and 90~au
spaced by 8 mutual Hill radii at the limits of long term stability
\citep{Chambers1996, Smith2009stability}. If we require planets to be
spaced by more than 8 mutual Hill radii, this multiplicity reduces
considerably to two or only one planet if planet masses are slightly
higher. We tested this by running new simulations with the same
parameters as in \S\ref{sec:nbody}, but instead with a pair of 30 or
10~\Me\ planets and semi-major axes ranging between 60 to 90~au. After
a few iterations we found a good fit using two 10~\Me\ planets with
semi-major axes of 60 and 83 au (i.e. spaced by 12 Mutual Hill
radii). Figure \ref{fig:radial_profile_nbody_2plt} shows the radial
profile for this model reproducing the width and depth of the gap. The
relative depth of the gap is similar to the one observed as there are
still particles on stable orbits at around 70 au and trapped in the
co-orbital regions of the two planets --- note that there are residuals
above $3\sigma$ at the inner and outer edge of the disc because this
model has sharp boundaries in semi-major axis. If we assumed that no
particles were present near the planet at the start of the simulation
as commented in \S\ref{sec:nbody}, then even lower mass planets
creating narrower gaps would be needed in order to achieve an overall
gap depth of 50\%.

\subsubsection{Planet(s) on eccentric orbits}

Although a single or multiple low mass planets might explain the
observed gap, the formation of an ice giant planet at $\sim80$~au
encounters significant difficulties compared to within a few tens of
au (see \S\ref{sec:pformation}). The scenario proposed by
\cite{Pearce2015doublering}, where a planet opens a broad gap through
secular interactions, circumvents some of these problems as the planet
is formed closer in at 10-20~au, and scattered out by a more massive
planet onto an eccentric orbit with a larger semi-major axis. This
scenario was able to fit the mean radial profile derived by
\cite{Ricci2015} and predicted the presence of asymmetries in the form
of spiral features and a small offset between the inner and outer
regions of the disc due to secular interactions with the
planet. However, no significant offset between the inner and outer
regions is present in our observations. By fitting an ellipse to the
inner and outer bright arcs in the band 6 image (roughly at 50 and
110~au), we constrain the offset to be $0\farcs05\pm0\farcs02$,
(i.e. consistent with zero and lower than 1.6~au). This translates to
a maximum eccentricity of 0.03 if we assume that the outer bright arc
is circular while the inner arc is eccentric and thus offset from the
star. Moreover, our observations suggest that the disc inner edge is
much steeper than predicted by \cite{Pearce2015doublering}, thus
inconsistent with their model.  A last recently proposed scenario
involves two interior planets with low but non-zero eccentricity,
which open a gap in the outer regions due to secular resonances
\citep[see][]{Yelvertoninprep}. Although this scenario can produce an
observable gap, in its simplest form the model produces a gap that is
narrower than seen for HD~107146.



\subsubsection{Dust-gas interactions}

There are other scenarios that might not require the presence of
planets to produce a gap or multiple ring structures in the dust
distribution. Photoelectric instability \citep{Klahr2005, Besla2007,
  Lyra2013Natur, Richert2017} is one of those scenarios and has
received particular attention lately as the presence of vast amounts
of gas around a few debris discs suggests that this mechanism might be
common. However, this mechanism is only important when dust-to-gas
ratios are comparable and it is not yet clear if relevant for the dust
distribution of large millimetre-sized grains that are not well
coupled to the gas. Assuming some residual primordial gas could still
be present in the disc, we convert our CO gas mass upper limit of
$5\times10^{-6}$~\Me\ (see \S\ref{dis:gas} below) to a total gas mass
upper limit of $\sim0.05$~\Me\ (assuming a CO/H$_2$ ratio of
$10^{-4}$) or a surface density of $\sim10^{-5}$~g~cm$^{-2}$. This
upper limit is comparable to the total dust mass in millimetre grains,
thus photoelectric instability could occur. However, using this gas
surface density upper limit we estimate a Stokes number of $10^{5}$
for millimetre grains, thus the stopping time is much longer than the
collisional lifetime of millimetre dust and of the order of the age of
the system. Therefore, we conclude that photoelectric instability does
not play an important role in the formation of the structure observed
by ALMA around HD~107146.

\subsection{Planet formation at tens of au}
\label{sec:pformation}

Based on these new ALMA observations, if the gap was cleared by a
single or multiple planets, these must have a low mass
($\lesssim$30~\Me) and formed between 50 and 100~au. Otherwise, if the
planets were formed closer in and scattered out onto an eccentric
orbit, the disc would appear asymmetric
\citep{Pearce2015doublering}. Such low planet masses and orbits at
tens of au resemble the ice giant planets in the Solar System, but
with a semi-major axis 2-3 times larger than Neptune's. \emph{Could
  such planets have formed in situ as these observations suggest?}
HD~107146's broad and massive debris disc indicates that planetesimals
efficiently formed at a large range of radii from 40 to 140
au. Moreover, the mass of these putative planets ($\lesssim30$~\Me) is
consistent with the solid mass available in their feeding zones
($\sim4$ Hill radii wide), based on the dust surface density derived
in \S\ref{sec:parmodel} and extrapolating it to the total mass surface
density (assuming a size distribution $dN\propto D^{-3.5}dD$ and
planetesimals up to sizes of 10 km). Nevertheless, in situ formation
encounters the two following problems. First, although a planet at
$\sim80$~au might grow through pebble accretion fast enough to form an
ice giant before gas dispersal \citep{Johansen2010, Ormel2010,
  Lambrechts2012, Morbidelli2012pa, Bitsch2015}, it requires the
previous formation of a massive planetesimal (or protoplanet) of
$10^{-2}-10^{-1}$~\Me\ (so-called transition mass). However, newly
born planetesimals through streaming instability \citep{Youdin2005}
have characteristic masses of rather
$10^{-6}-10^{-4}$~\Me\ \citep{Johansen2015, Simon2016}. These
planetesimals can grow through the accretion of pebbles and smaller
planetesimals, but this growth (so-called Bondi accretion) is very
slow for low mass bodies at large stellocentric distances
\citep{Johansen2015}. A possible solution to this problem is that the
protoplanetary disc around HD~107146 was unusually massive and
long-lived, which would increase the chances of forming such a
protoplanet. Alternatively, the low mass protoplanet could have formed
closer in and been scattered out and circularised during the
protoplanetary disc phase. The second problem that in-situ formation
faces is related to planet migration. A protoplanet growing to form an
ice giant is expected to migrate inwards through type-I migration
\citep{Tanaka2002}. While this might imply that the single or multiple
putative planets might have formed at larger radii, our observations
suggest that they would have attained most of their mass within the
observed gap, i.e. within 100~au. If pebble accretion is fast enough,
the planet could grow fast and migrate only slightly before disc
dispersal.

\emph{Why would planets form only between 60-90~au within this 100 au
  wide disc of planetesimals?} While no planets might have formed
beyond 90~au as planetesimal accretion was too slow for a planetesimal
to reach the transition mass and efficiently accrete pebbles, this is
not the case for planetesimals formed at smaller radii between the
inner edge of the disc and the orbit of the innermost putative
planet. \textit{Why did only planetesimals form between 40-60~au?}
Planetesimal growth might have been hindered there if the orbits of
large solids were stirred by inner planets, making their accretion
rate slower. Alternatively, planetesimal growth between 60-90~au could
have been more efficient due to a local enhancement in the available
solid mass. As stated before, it is also possible that the low mass
protoplanet did not form in situ, but it was scattered out from
further in.

\subsection{Massive planetesimal disc}
\label{sec:massparticles}

Here we discuss the effect that a massive planetesimal disc could have
on the conclusions stated above where we assumed a disc of negligible
mass ($\ll10$~\Me). As shown in \S\ref{sec:model}, this debris disc is
likely very massive and thus affects the dynamics of this system,
e.g. because of the gravitational force on the planet and disc
self-gravity. A massive disc can induce planetesimal driven migration
where the planet migrates through a planetesimal disc due to the
angular momentum exchange in close encounters
\citep[e.g.][]{Fernandez1984, Ida2000, Gomes2004}.
This type of migration has been well studied in the context of the
outer Solar System, as it could have driven an initially compact
orbital configuration to a more extended and current configuration
\citep{Hahn1999, Hahn2005}, or towards an orbital instability
\citep[e.g.][]{Tsiganis2005}. In all these models the outer planets
scatter material in to Jupiter which ejects most of it, leading to an
outward migration of the outer planets and a small inward migration of
Jupiter. However, \cite{Ida2000} showed that even in the absence of
interior planets self-sustained migration could have led Neptune's
orbit to expand due to the asymmetric planetesimal distribution around
it. A more recent work by \cite{Kirsh2009}, however, showed that a
single planet embedded in a planetesimal disc migrates preferentially
inwards due to the timescale difference between the inner and outer
feeding zones \citep[see also][]{Ormel2012}. Therefore, we expect that
the putative planet at $\sim$80~au around HD~107146 is likely
migrating or has migrated inwards since it formed.

Simple scaling relations from \cite{Ida2000} predict that the
migration rate should be of the order of
\begin{equation}
  \left| \frac{da}{dt} \right| \approx \frac{4\pi\Sigma a^2}{M_{\star}}\frac{a}{T} \label{eq:migration}
\end{equation}
which agrees roughly with numerical simulations
\citep[e.g.][]{Kirsh2009}. Given the expected surface density of
planetesimals at 80~au around HD~107146
($\Sigma\gtrsim10^{-2}$~\Me~au$^{-2}$ for a maximum planetesimal size
of 10 km), Equation \ref{eq:migration} predicts that the migration
rate is such that the planet would have crossed the whole disc
reaching its inner edge in only a few Myr. \cite{Kirsh2009} found,
however, that when the planet mass exceeds that of the planetesimals
within a few Hill radii, the migration rate decreases strongly with
planet mass. Assuming that the gap is indeed caused by a single or
multiple planets between 60-90~au, given the 30~\Me\ upper limit for
the planet mass we conclude that the surface density of planetesimals
must be much lower than 0.01~\Me~au$^{-2}$ to hinder planetesimal
driven migration, i.e. a total disc mass $\lesssim430$~\Me. If not the
surface density profile would probably be significantly different
without a well defined gap. This dynamical upper limit together with
the lower limit derived from collisional models constrain the total
disc mass to be between $\sim300-400$~\Me, which is close to the
maximum solid mass available in a protoplanetary disc under standard
assumptions (e.g. disc-to-stellar mass ratio of 0.1 and gas-to-dust
mass ratio of 100). This particularly high disc mass at the limits of
feasibility is not unique, but a confirmation of the so-called ``disc
mass problem'' as many other young and bright discs need similar or
even higher masses according to collisional evolution models
\citep[see discussion in][]{Krivov2018}.

Although we expect that the gap produced by a migrating planet might
look significantly different compared with the no migration scenario
(e.g. wider and radially and perhaps azimuthally asymmetric), none of
the above studies provided a prediction for the resulting surface
density of particles during planetesimal driven migration that we
could compare with our observations. In experimental runs with a 10 or
30~\Me\ planet and a similar mass planetesimal disc (using massive
test particles) we find that the planet migrates inward $\sim$5-40~au
in 100~Myr depending on the planet and disc mass, and that the gap has
an asymmetric radial profile that is significantly distinct from the
no-migration scenario, which could explain why we find a slight
asymmetry within the gap (see \S\ref{sec:continuum}). Future
comparison with numerical simulations of planetesimal-driven migration
could provide evidence for inward or outward migration, and thus set
tighter constraints on the disc and planet mass. Dynamical estimates
of the disc mass could also shed light on the disc mass problem,
confirming the high mass derived from collisional models or rather
indicating that these need to be revisited.

\subsection{Warm inner dust component}
\label{dis:warmdust}

While IR excess detections can be generally explained by a single
temperature black body, there is a large number of systems that show
evidence for a broad range of temperatures that are hard to explain
simply as due to a single dusty narrow belt, even with temperatures
varying as a function of grain size \citep[e.g.][]{Backman2009,
  Morales2009, Chen2009, Ballering2014, Kennedy2014}. Although this
might be explained by material distributed over a broad range of radii
(like in protoplanetary discs), an alternative and very attractive
explanation for these systems is the presence of a two-temperature
disc, with an inner asteroid belt and an outer exo-Kuiper belt,
analogous to the Solar System \citep[e.g.][]{Kennedy2014,
  Schuppler2016, Geiler2017}. HD~107146 is a good example of this type
of system \citep{Ertel2011, Morales2011}, with a significant excess at
22~$\mu$m that cannot be reproduced by models of a single outer belt,
but rather is indicative of dust located within $\sim30$~au produced
in an asteroid belt. These type of belts are typically hard to resolve
due to the small separation to the star (hindering scattered light
observations) where current instruments are not able to resolve and
due to their low emissivity that peaks in mid-IR.

\begin{figure}
  \centering
 \includegraphics[trim=0.0cm 0.0cm 0.0cm 0.0cm, clip=true, width=1.0\columnwidth]{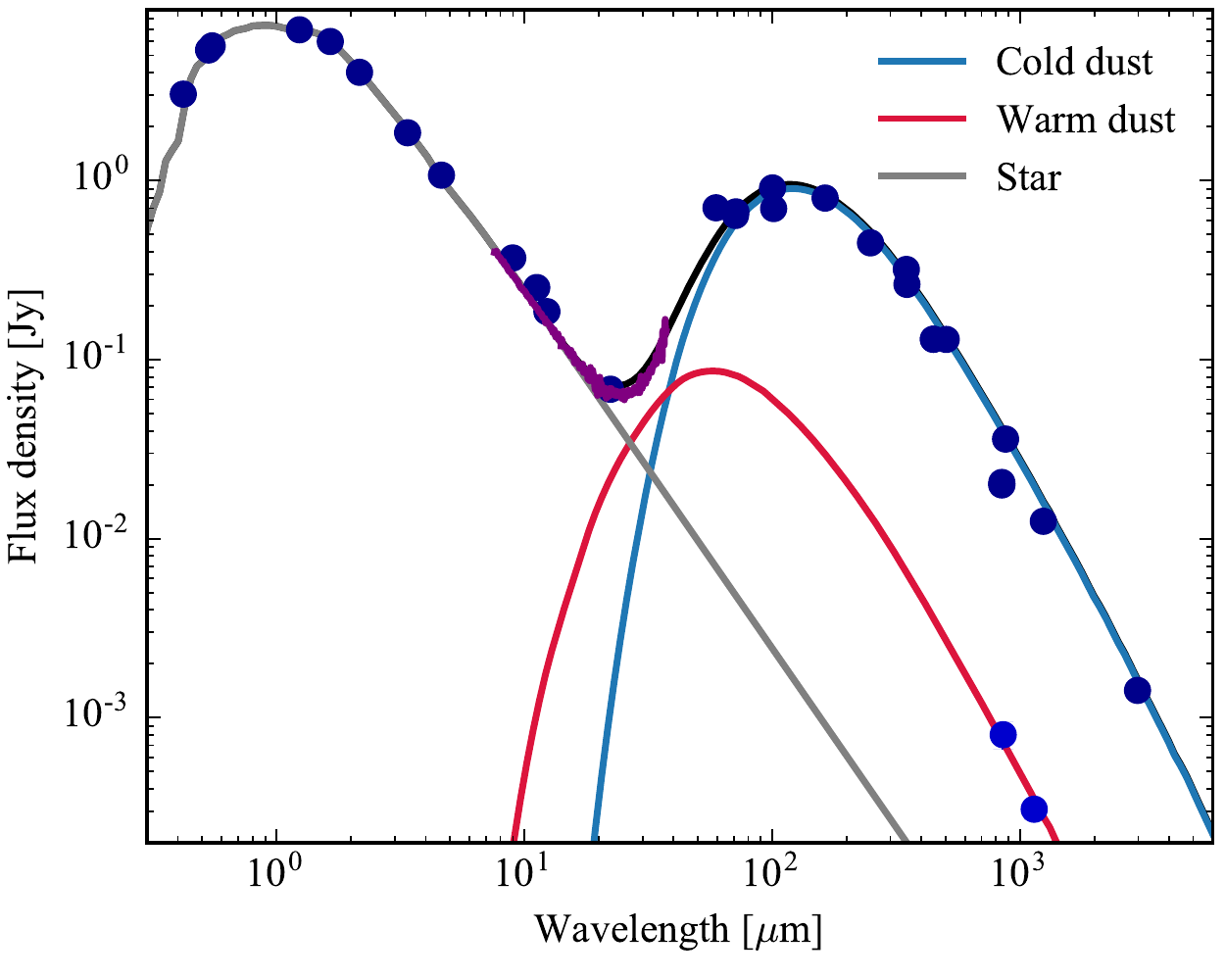}
 \caption{Spectral Energy Distribution of HD~107146 \citep[dark blue
     points,][]{Kennedy2014} and its inner component (light blue
   colours) obtained after subtracting the outer component of our
   model presented in \S\ref{sec:modelin}. The grey, light blue and
   red lines represent the stellar, outer disc, inner component
   contribution to the total flux (black line), respectively.} 
 \label{fig:sed}
\end{figure}

\emph{Is the detected inner emission related to the $\sim20$~$\mu$m
  excess?} In order to test if the inner emission seen by ALMA is
compatible with being warm dust, we use the parametric model developed
in \S\ref{sec:parmodel} to see if it can reproduce the available
photometry of this system, including the mid-IR excess. We introduce,
though, two changes to the model assumptions: first, we modify the
size distribution index from -3.5 to -3.36 to be consistent with the
derived spectral index in this work and previous studies
\citep{Ricci2015atca}; and second, we extend the size distribution
from 1 to 5~cm. The latter is necessary to reproduce the photometry at
7~mm as the contribution from cm-sized grains is significant at these
wavelengths. In Figure \ref{fig:sed} we compare the model and observed
SED, including the new ALMA photometric points of the inner
component. Despite the simplicity of our model, it reproduces
successfully both the photometry at 22 $\mu$m and at millimetre
wavelengths, confirming that our detection is consistent with being
warm dust emission.

\begin{figure}
  \centering
 \includegraphics[trim=0.1cm 0.5cm 0.2cm 0.5cm, clip=true, width=1.0\columnwidth]{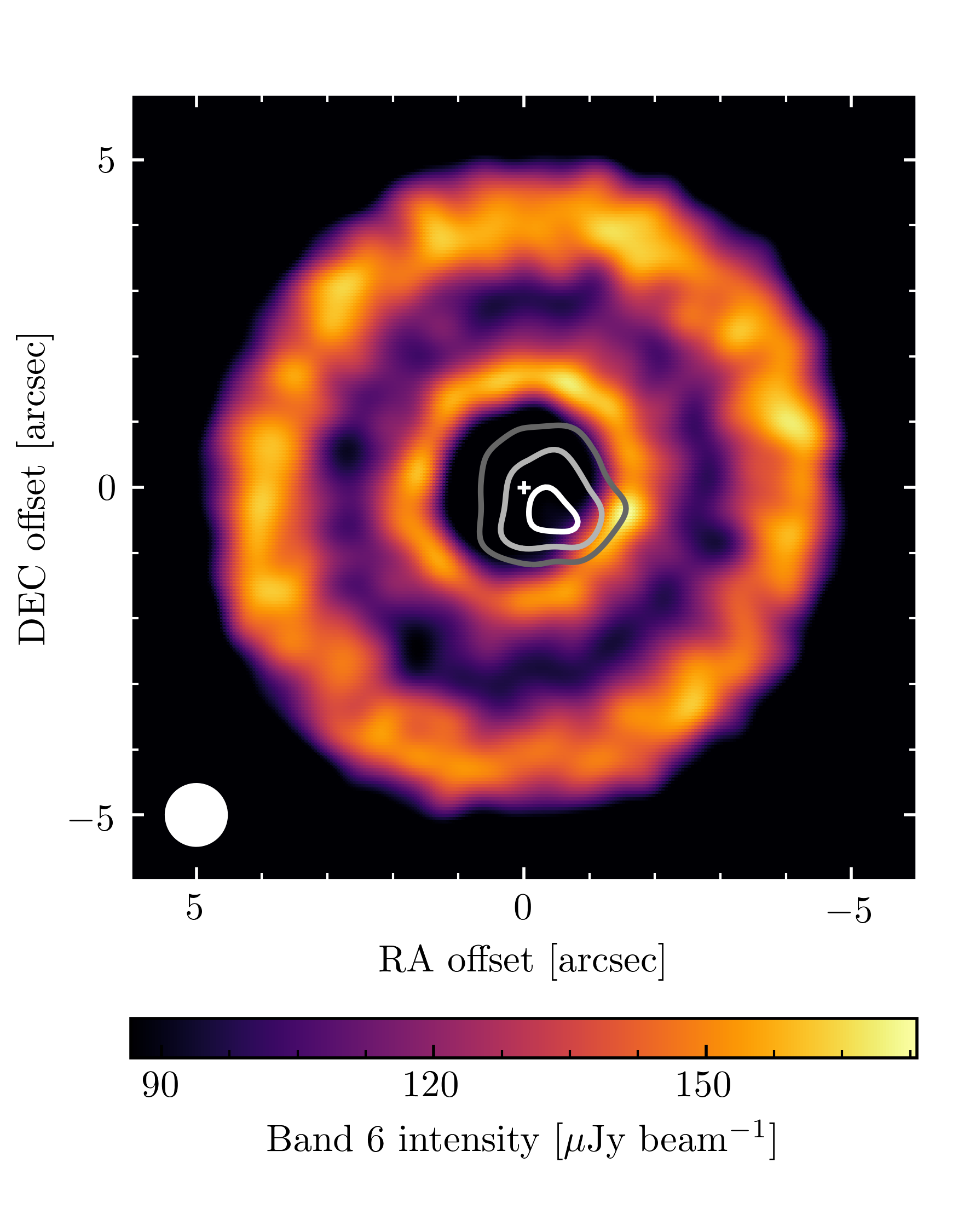}
 \caption{Clean image of HD~107146 at 1.1 mm (band 6) using natural
   weights. Overlayed in contours are the residuals after subtracting
   the best fit model presented in \S\ref{sec:modelin}, but without
   the inner component. Contour levels are set to 4, 8 and 12
   $\sigma$. The stellar position is marked with a white cross at the
   center of the image, while the beam ($0\farcs80\times0\farcs79$) is
   represented by a white ellipse in the bottom left corner. The image
   rms at the center is 6.3 $\mu$Jy~beam$^{-1}$. For a better display
   we have adjusted the color scale with a minimum of 50\% the peak
   flux.}
 \label{fig:continuum_clump}
\end{figure}

These new ALMA observations provide unique information on the nature
of the warm dust emission as it is marginally resolved (see Figure
\ref{fig:continuum_clump}). The surface brightness peaks at
$\sim20$~au, thus roughly in agreement with the predicted location
based on SED modelling. The emission, however, is far from originating
in an axisymmetric asteroid belt, but is rather asymmetric with a
maximum brightness towards the south west side of the disc. \emph{What
  could cause such an asymmetric dust distribution?} We identify three
scenarios proposed in the literature that could cause long- or
short-term brightness asymmetries in a disc or belt.

First, the inner emission could correspond to an eccentric disc, which
at millimetre wavelengths would be seen brightest at apocentre. This
is known as apocentre glow, which is caused by the increase in dust
densities at apocentre for a coherent disc \citep{Wyatt2005secular,
  Pan2016}. Disc eccentricities can be caused by perturbing planets
\citep[e.g.][]{Wyatt1999, Nesvold2013, Pearce2014}, as it has been
suggested to explain Fomalhaut's eccentric debris disc
\citep{Quillen2006, Chiang2009, Kalas2013, Acke2012, MacGregor2017}
and HD~202628 \citep{Krist2012, Thilliez2016, Faramazinprep}. At long
wavelengths, the contrast between apocentre and pericentre brightness
is expected to be approximately \citep{Pan2016}
\begin{equation*}
  \left(\frac{1-e/2}{1+e/2}\right)\left(\frac{1+e}{1-e}\right),
\end{equation*}
thus to reach a contrast higher than 2 (as observed for HD~107146's
inner disc), the disc eccentricity would need to be higher than
0.5. Assuming the disc eccentricity corresponds to the forced
eccentricity (i.e. free eccentricities are much smaller), then the
perturbing planet should have a very high eccentricity
$\gtrsim0.5$. One potential problem with this scenario is that the
outer disc does not show any hint of being influenced by an eccentric
planet. This problem could be circumvented if the true eccentricity of
the inner belt is lower than derived based on the image residuals
(e.g. $\lesssim0.2$) and the inner planet has a semi-major axis of
only a few au, as the forced eccentricity on the outer disc would be
much smaller.  Alternatively, if the disc or putative outer planet(s)
are much more massive than the inner eccentric planet, the disc might
remain circular as it is observed.

A second potential scenario relates to a recent collision between
planetary embryos, which would release large amounts of dust at the
collision point producing an asymmetric dust distribution that could
last for $\sim1000$ orbits or $\sim1$~Myr at 20~au \citep{Kral2013,
  Jackson2014}. In such a scenario the \emph{pinch point} would appear
brightest since the orbits of the generated debris converge where the
impact occurred and more debris is created from collisions. Thus, the
pinch point would appear radially narrow. However, both our band 6 and
7 datasets show that the emission is radially broad at its brightest
point, spanning $\gtrsim20$~au. This could be circumvented if the
collision occurred within a broad axisymmetric disc. Higher resolution
observations are necessary to discard this scenario or confirm these
two inner components.

Finally, a third possible scenario is that the asymmetric structure is
caused by planetesimals trapped in mean motion resonances (typically
3:2 and 2:1) with an interior planet that migrated through the
planetesimal disc \citep{Wyatt2002, Wyatt2003, Wyatt2006,
  Reche2008}. The exact dust spatial distribution depends on the
planet mass, migration rate and eccentricity. The single clump
inferred from our observations suggests that planetesimals would be
trapped predominantly on the 2:1 resonance rather than 3:2 as the
latter only creates a two clump symmetric structure. Simulations by
\cite{Reche2008} showed that in order to trap planetesimals in the 2:1
resonance, a Saturn mass planet (or higher) with a very low
eccentricity was needed, therefore, placing a lower limit on the
planet mass if the asymmetry is due to resonant trapping.

Future observations could readily distinguish between scenarios 1-2
and 3, as in the first two scenarios the orientation of the asymmetry
should stay constant (precession timescales are orders of magnitude
longer than the orbital period), while in the third scenario it should
rotate at the same rate as the putative inner planet orbits the
star. Moreover, higher resolution and more sensitive observations
could reveal if the disc is eccentric, smooth with a radially narrow
clump, or smooth with a radially broad clump (scenarios 1 to 3,
respectively). Observations at shorter wavelength would also be
useful. In the eccentric disc scenario we expect the disc to be
eccentric and broader due to radiation pressure on small dust grains,
while resonant structure would be completely absent and the disc
should look axisymmetric as small grains are not trapped also due to
radiation pressure. Finally, future ALMA observations should be able
to definitely rule out the possibility of the inner emission arising
from a sub-mm galaxy. Given HD~107146's proper motion (-174 and -148
mas~yr$^{-1}$ in RA and Dec. direction, respectively), we would expect
that in 2 years any background object should shift by $0\farcs23$
towards the north east with respect to HD~107146, thus enough to be
measured with ALMA observations of similar sensitivity and resolution
to the ones presented here.

\subsection{Gas non detections}
\label{dis:gas}

\begin{figure}
  \centering
 \includegraphics[trim=0.1cm 0.3cm 0.0cm 0.3cm, clip=true, width=1.0\columnwidth]{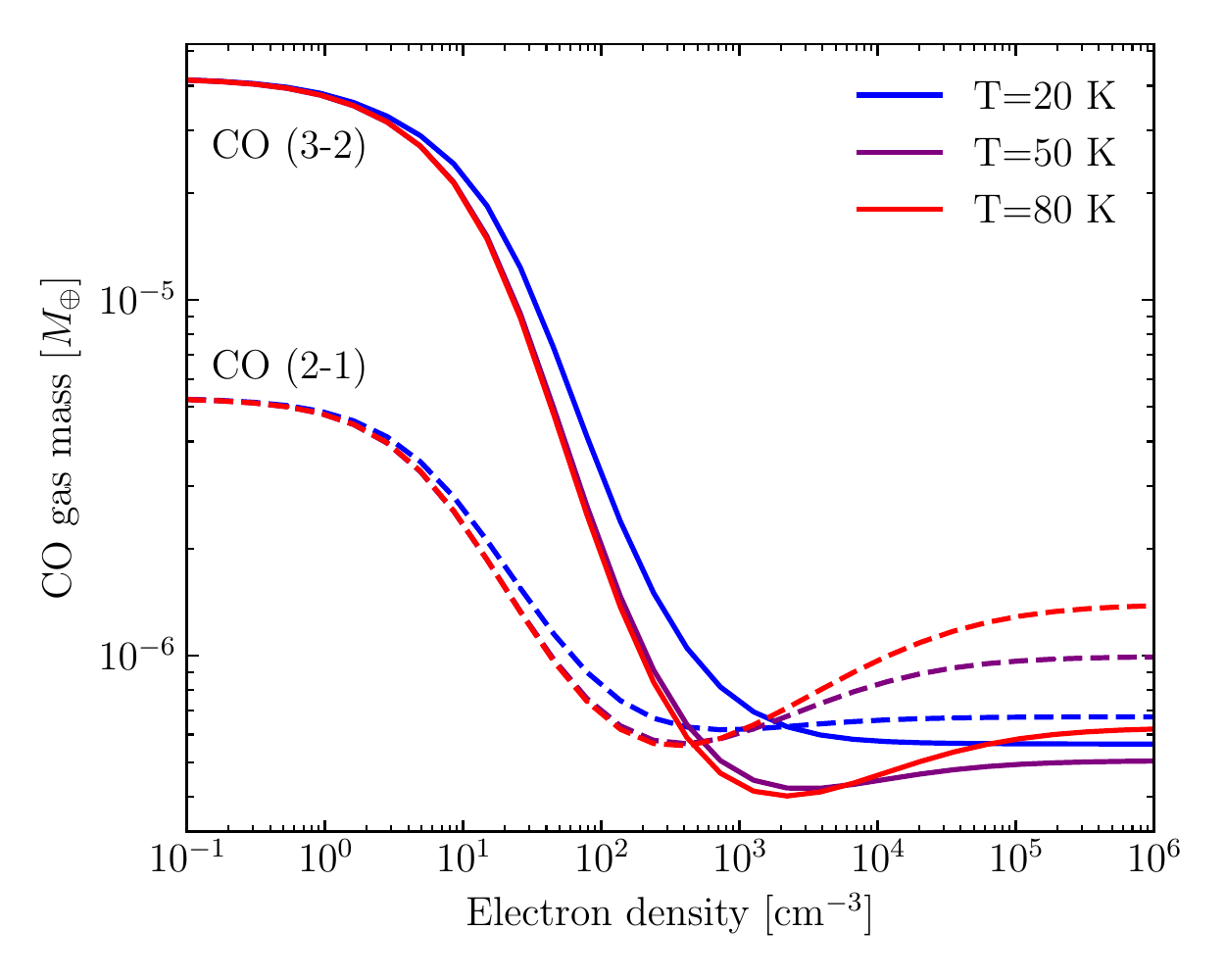}
 \caption{CO gas mass upper limit based on CO J=3-2 (continuous line)
   and 2-1 (dashed line) flux upper limits, and as a function of
   electron densities and gas kinetic temperature. The blue, purple
   and red lines represent gas kinetic temperatures of 20, 50 and
   80~K, respectively.}
 \label{fig:COmass}
\end{figure}

In \S\ref{sec:gas} we search for CO and HCN secondary origin gas
around HD~107146. Although we did not find any, here we use the flux
upper limits, including also the 40 mJy~km~s$^{-1}$ upper limit for CO
J=2-1 from \cite{Ricci2015}, to derive an exocometary gas mass upper
limit in this disc. It has been demonstrated that in the low density
environments around debris discs, gas species are not necessarily in
local thermal equilibrium (LTE), which typically leads to an
underestimation of CO gas masses \citep{Matra2015}. Here we use the
code developed by \citep{Matra2018} to estimate the population of the
CO rotational level in non-LTE based on the radiation environment,
densities of collisional partners, and also taking into account UV
fluorescence. We consider as sources of radiation the star, the CMB
and the mean intensity due to dust thermal emission within the disc
(calculated using our model and RADMC-3D). We choose 80~au as the
representative radius, which is approximately the middle radius of the
disc, and left as a free parameter the density of collisional partners
(here assumed to be electrons). Based on our model, the dust
temperature should roughly vary between 50 to 30~K between the disc
inner and outer edge. Figure \ref{fig:COmass} shows the CO gas mass
upper limit as a function of electron density for different kinetic
temperatures. We find that for electron densities lower than
$\sim10^2$~cm$^{-3}$, the CO J=2-1 upper limit is more constraining
than J=3-2 and vice versa. Overall, the CO gas mass must be lower than
$5\times10^{-6}$~\Me. We then use equation 2 in
\cite{Matra2017fomalhaut} and the estimated photodissociation
timescale of 120 yr to estimate an upper limit on the CO+CO$_{2}$ mass
fraction of planetesimals in the disc, but we find no meaningful
constraint because this CO gas mass upper limit is much higher than
the predicted $5\times10^{-7}$~\Me\ if gas were released in collisions
of comet-like bodies \citep[e.g.][]{Marino2016, Kral2017CO,
  Matra2017betapic, Matra2017fomalhaut}.

In the absence of a tool to calculate the population of rotational
levels for HCN, we estimate a mass upper limit assuming LTE. For
temperatures ranging between 30-50~K, we find an upper limit of
$3\times10^{-9}$~\Me, which translates to an upper limit on the mass
fraction of HCN in planetesimals of 3\%, an order of magnitude higher
than the observed abundance in Solar System comets
\citep{Mumma2011}. Note that this HCN upper limit could be much higher
due to non-LTE effects, thus this 3\% limit must be taken with
caution.

\section{Conclusions}
\label{sec:conclusions}

In this work, we have analysed new ALMA observations of HD~107146's
debris disc at 1.1 and 0.86 mm, to study a possible planet-induced gap
suggested by \cite{Ricci2015} with a higher resolution and
sensitivity. These new observations show that HD~107146, a 80-200~Myr
old G2V star, is surrounded by a broad disc of planetesimals from 40
to 140~au, that is divided by a gap $\sim$40~au wide (FWHM), centered
at 80~au and 50\% deep, i.e. the gap is not devoid of material. We
constrained the disc morphology, mass and spectral index by fitting
parametric models to the observed visibilities using an MCMC
procedure. We find that the disc is consistent with being
axisymmetric, and we constrain the disc eccentricity to be lower than
0.03.

The observed morphology of HD~107146's debris disc suggests the
presence of a planet on a wide circular orbit opening a gap in a
planetesimal disc through scattering. We run a set of N-body
simulations of a planet embedded in a planetesimal disc that we
compare with our observations. We find that the observed morphology is
best fit with a planet mass of 30~\Me, but significant residuals
appear after subtracting the best fit model. We conclude that the
observed gap cannot be reproduced by the dynamical clearing of such a
planet as the gap it creates is significantly deeper and narrower than
observed. We discuss that this could be circumvented by allowing the
planet to migrate (e.g. due to planetesimal driven migration) or by
allowing multiple planets to be present.  We discuss how a planet
could have formed in situ if the primordial disc was massive and
long-lived, and possibly grew to its final mass very quickly and by
the end of the disc lifetime (e.g. through pebble accretion), avoiding
significant inward migration and runaway gas accretion. Moreover,
because the putative planet(s) could undergo very fast planetesimal
driven migration, we set an upper limit on the surface density and
total mass of the disc.


These ALMA observations also revealed unexpected emission near the
star that is best seen when subtracting our best fit parametric model
of the outer disc. This inner component has a total flux of 0.8 and
0.3 mJy at 0.86~mm and 1.1~mm, respectively, a peak intensity that is
significantly offset from the star by $0\farcs5$ (15~au), and we
resolve the emission both radially and azimuthally. Its radial
location indicates that it could be the same warm dust that had been
inferred to be between 10-15~au to explain HD~107146's excess at
$22\mu$m. Indeed, we fit this emission with an extra inner asymmetric
component finding a good match with these ALMA observations and also
with HD~107146's excess at $22\mu$m. We constrain its peak density at
19~au, and a radial width of at least 20~au. We hypothesise that this
asymmetric emission could be due to a disc that is eccentric due to
interactions with an eccentric inner planet, asymmetric due to a
recent giant collision, or clumpy due to resonance trapping with a
migrating inner planet. On the other hand, we find that this inner
emission is unlikely to be a background sub-mm galaxy, as the
probability of finding one as bright as 0.8 mJy at 0.86~mm within the
disc inner edge (i.e. co-located with the warm dust) is 0.1\%.

Finally, although it had been demonstrated by \cite{Ricci2015} that no
primordial gas is present around HD~107146, we search for CO and HCN
gas that could be released from volatile-rich solids throughout the
collisional cascade in the outer disc. However, we find no gas, but we
place upper limits on the total gas mass and HCN abundance inside
planetesimals, being consistent with comet-like composition.

\section*{Acknowledgements}

We thank Bertram Bitsch for useful discussion on planet formation and
pebble accretion. JC and VG acknowledge support from the National
Aeronautics and Space Administration under grant No. 15XRP15\_20140
issued through the Exoplanets Research Program. MB acknowledges
support from the Deutsche Forschungsgemeinschaft (DFG) through project
Kr 2164/15-1. VF's postdoctoral fellowship is supported by the
Exoplanet Science Initiative at the Jet Propulsion Laboratory,
California Inst. of Technology, under a contract with the National
Aeronautics and Space Administration. GMK is supported by the Royal
Society as a Royal Society University Research Fellow. LM acknowledges
support from the Smithsonian Institution as a Submillimeter Array
(SMA) Fellow. This paper makes use of the following ALMA data:
ADS/JAO.ALMA\#2016.1.00104.S and 2016.1.00195.S. ALMA is a partnership
of ESO (representing its member states), NSF (USA) and NINS (Japan),
together with NRC (Canada) and NSC and ASIAA (Taiwan) and KASI
(Republic of Korea), in cooperation with the Republic of Chile. The
Joint ALMA Observatory is operated by ESO, AUI/NRAO and
NAOJ. Simulations in this paper made use of the REBOUND code which can
be downloaded freely at http://github.com/hannorein/rebound.




\bibliographystyle{mnras}
\bibliography{SM_pformation} 



\appendix

\section{Linear scaling of semi-major axis}
\label{appendix}

\begin{figure}
  \centering
 \includegraphics[trim=0.3cm 0.3cm 0.3cm 0.3cm, clip=true, width=1.0\columnwidth]{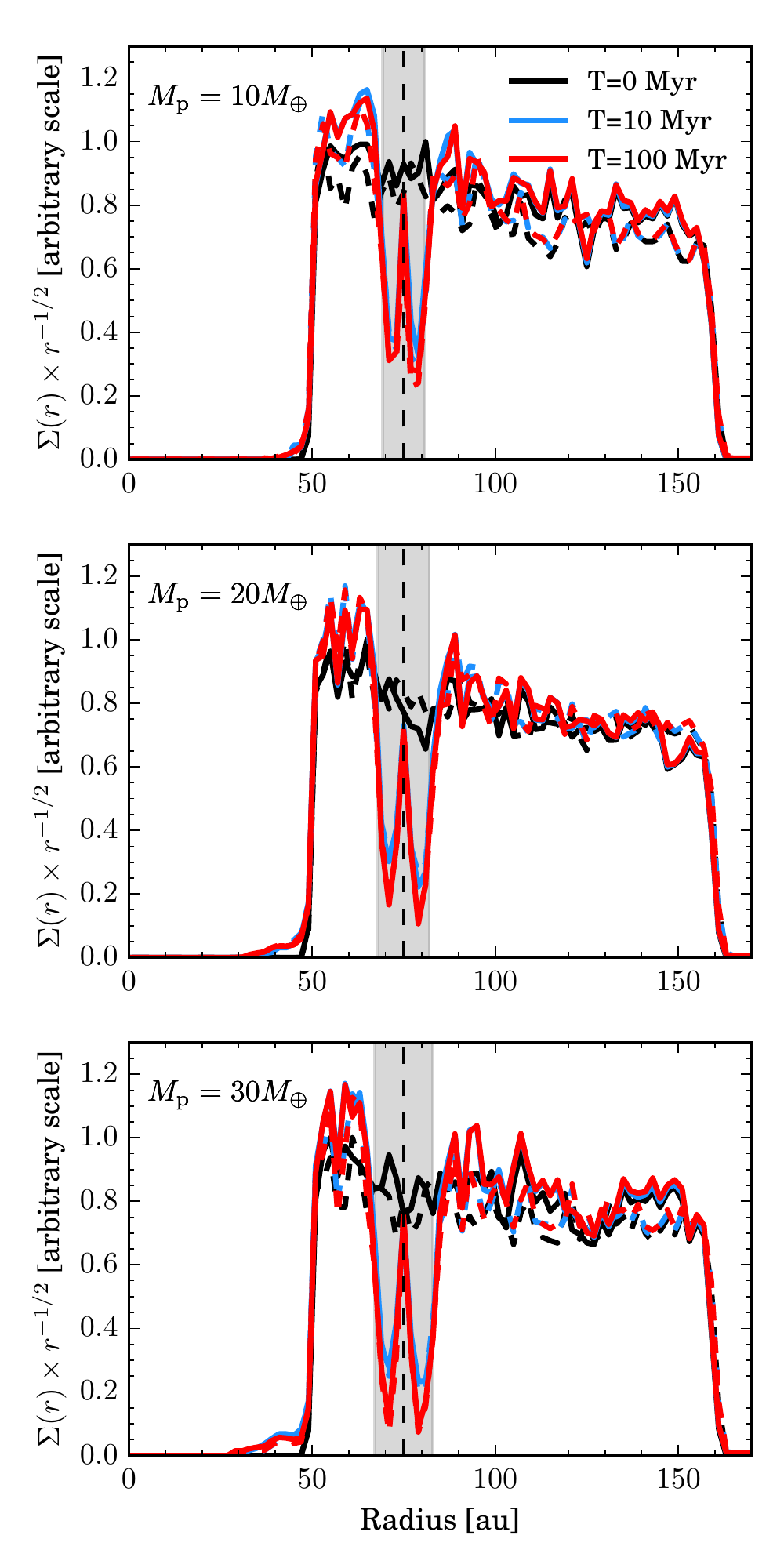}
 \caption{Surface brightness evolution of disc perturbed by a planet
   on a circular orbit at 75~au. The continuous and dashed lines
   represent simulations run with a planet with a semi-major axis of
   80~au and then scaled to 75~au, and with a planet with a semi-major
   axis of 75~au without scaling the simulation, respectively. Surface
   brightness of particles assuming an initial surface density
   proportional to $r^{0.25}$ and a dust temperature profile
   decreasing with radius as $r^{-1/2}$. The grey shaded region
   represents the chaotic zone approximated by
   $\ap\pm2\ap(\Mp/M_{\star})^{2/7}$. The top, middle and bottom
   panels show systems with planet masses of 10, 20 and 30~\Me,
   respectively.}
 \label{fig:nbody_check}
\end{figure}

In \S\ref{sec:nbody} we assumed that we can approximate the surface
density of particles interacting with a planet at a given semi-major
axis (between $\sim75-80$~au), by linearly scaling the distances in
the output of a N-body simulation of a planet on a fixed semi-major
axis of 80~au. In order to check the validity of this, we run three
additional simulations with a planet semi-major axis of 75~au and
masses of 10, 20 and 30~\Me. Figure \ref{fig:nbody_check} compares
the resulting surface density of these new simulations (dashed line)
with simulations of a planet at 80~au, but scaled to 75~au. These
three new simulations show that the linear scaling is a reasonable
approximation, matching very well the structure seen in the
simulations with a planet at 75~au. Only minor differences are
present, which are of the order of the expected noise due to the
random initial semi-major axis of particles in our simulations.

\bsp	
\label{lastpage}
\end{document}